\documentclass[twocolumn]{aastex61}

\usepackage{xspace,graphics,graphicx,longtable,amsmath}
\usepackage{natbib,amsbsy}
\usepackage{appendix}
\def \lya {Ly$\alpha$ }
\def \mkms {{\rm \; km\;s^{-1}}}

\def \mrperp {R_{\perp}}
\defcitealias{GPG1}{GPG1}

\newcommand{\msun}{~M_{\odot}}
\newcommand{\overbar}[1]{\mkern 1.5mu\overline{\mkern-1.5mu#1\mkern-1.5mu}\mkern 1.5mu}
\providecommand{\cllength}{\ensuremath{1.9}}
\providecommand{\clarea}{\ensuremath{3.5}}

\shorttitle{Galaxies Probing Galaxies in PRIMUS -- II.}
\shortauthors{Rubin et al.}
\accepted{July 21, 2018}

\begin{document}

\title{Galaxies Probing Galaxies in PRIMUS -- II. The Coherence Scale
  of the Cool Circumgalactic Medium}

\author{Kate H. R. Rubin}
\affiliation{San Diego State University, Department of Astronomy,
  San Diego, CA 92182, USA} 

\author{Aleksandar M. Diamond-Stanic}
\affiliation{Bates College, Department of Physics \& Astronomy, 44
Campus Ave, Carnegie Science Hall, Lewiston, ME 04240, USA}, 

\author{Alison L. Coil}
\affiliation{Center for Astrophysics and Space Sciences,
  Department of Physics, University of California, San Diego, 9500 Gilman Drive, La Jolla, CA 92093, USA}

\author{Neil H. M. Crighton}
\affiliation{Center for Astrophysics and Supercomputing,
  Swinburne University of Technology, Hawthorn, VIC 3122, Australia}

\author{Kyle R. Stewart}
\affiliation{Department of Mathematical Sciences, California Baptist University, 8432 Magnolia Ave., Riverside, CA 92504, USA}

\correspondingauthor{Kate H. R. Rubin}
\email{krubin@sdsu.edu}

\begin{abstract}
The circumgalactic medium (CGM) close to ${\sim}L_*$ star-forming galaxies
hosts strong \ion{Mg}{2} $\lambda2796$ absorption (with equivalent width $W_{2796}>0.1~\rm\AA$)
with a near-unity covering fraction.  To characterize the spatial coherence of this absorption,
we analyze the $W_{2796}$ distribution in the CGM of $27$ star-forming galaxies
detected in deep spectroscopy of bright background (b/g) galaxies first presented in \citet{GPG1}.
The sample foreground (f/g) systems have redshifts $0.35\lesssim
z\lesssim0.8$ and stellar masses $9.1<\log M_*/M_{\odot}<11.1$, and the b/g galaxies provide spatially-extended probes
with half-light radii $1.0~\mathrm{kpc}\lesssim R_{\rm eff}\lesssim
7.9~\mathrm{kpc}$ at projected distances $\mrperp<50~\mathrm{kpc}$.  Our analysis also draws 
on literature $W_{2796}$ values measured in b/g
QSO spectroscopy probing the halos of f/g galaxies with a
similar range in $M_*$ at $z\sim0.25$.
By making the assumptions that 
(1) samples of like galaxies exhibit similar circumgalactic $W_{2796}$ distributions;
and that (2)  the quantity
$\log W_{2796}$ has a Gaussian distribution with a dispersion that is
constant with $M_*$ and $\mrperp$, we use this QSO-galaxy
pair sample to construct a model for the $\log W_{2796}$ distribution in the CGM of
low-redshift galaxies.  Adopting this model, we then demonstrate the dependence of the observed $\log W_{2796}$
distribution on the ratio of the surface area of the b/g probe to the
projected absorber surface area ($x_{\rm A}\equiv A_{\rm  G}/A_{\rm A}$), finding that
distributions which assume $x_{\rm A}\ge15$ are statistically inconsistent with that observed toward our b/g
galaxy sample at a $95\%$ confidence level.  This limit, in combination with the b/g galaxy sizes, 
requires that the length scale over which $W_{2796}$ does not vary (i.e., the ``coherence scale'' of \ion{Mg}{2} absorption) is
 $\ell_{\rm A}>\cllength~\mathrm{kpc}$.  This novel constraint
 on the morphology of cool, photoionized structures in the inner CGM suggests that either these structures each
 extend over kiloparsec scales, or that the numbers and velocity
 dispersion of these structures are spatially correlated over the same
 scales.  

\end{abstract}

\section{Introduction}\label{sec.intro}

Within the last decade, QSO absorption line experiments have revealed
the gaseous material enveloping low-redshift galaxies to be a dominant
component of their host halo's baryonic mass.  Galvanized by the
unprecedented sensitivity of the Cosmic Origins Spectrograph on the {\it
  Hubble Space Telescope}, these studies report a mass in cool (temperature $T\sim 10^4$
K), diffuse baryons of nearly $10^{11}\msun$ filling the regions
extending to $160$ kpc from isolated, ${\sim}L^*$ galaxies
\citep{Stocke2013,Werk2014,Peeples2014,Prochaska2017}.
Simultaneous observations of the
highly-ionized metal species \ion{O}{6} 
have been interpreted to indicate
the presence of another, warmer gas phase at 
$T\gtrsim10^5$ K, estimated to contain a mass of more than $10^9\msun$
\citep{Tumlinson2011,Prochaska2017}. 
Moreover, an enduring prediction of
galaxy formation theory is the shock-heating of gas as it falls onto
dark matter halos 
\citep{ReesOstriker1977,WhiteRees1978,Keres2005,Nelson2013},
resulting in a ubiquitous ``hot'' phase
($T \sim 10^6$ K) filling halos with masses
$\gtrsim 10^{11} \msun$ 
\citep{BirnboimDekel2003,Keres2009}.
These empirical and theoretical findings imply the omnipresence of a
massive gas reservoir composed of material over a broad range of
temperatures surrounding luminous galaxies in the nearby universe.

The predominance of this baryonic component in turn implies a crucial
role in the regulation of galaxy growth.  Indeed, hydrodynamical
simulations of galaxy formation predict that this reservoir is fed
 by the accretion of material from the intergalactic medium, by the
 stripping of satellite galaxies as they merge with the central
 massive host, and by large-scale outflows of gas driven from
 star-forming regions via feedback processes
 \citep[e.g.,][]{Oppenheimer2010,Shen2013,Ford2014,Hopkins2014}.
At the same time, these simulations predict distinct spatial
distributions and morphologies for each gas phase.
\ion{O}{6}-absorbing material, for instance, typically exhibits a
relatively smooth morphology extending well beyond the virial radius
of a halo of mass $M_h \sim10^{12}\msun$
\citep[e.g.,][]{Shen2013,Oppenheimer2016}, while optically thick
\ion{H}{1} and gas traced by 
absorption in low-ionization metal transitions (e.g., \ion{Si}{2},
\ion{C}{2}) is distributed in narrow filaments or small clumps
\citep{Shen2013,Fumagalli2014,Faucher-Giguere2015}.  These morphologies may ultimately be linked to the
physical origin of each phase, and therefore may potentially  
corroborate interpretations based on other factors (e.g.,
metallicity or kinematics).



Moreover,
a characterization of the detailed structure of circumgalactic
material is crucial to our understanding of its hydrodynamics
\citep{Crighton2015}.
As the region through which gas accretes
onto galaxies, and as the reservoir 
receiving galactic wind ejecta, 
the circumgalactic medium (CGM) cannot be understood as a static gaseous
body
\citep[e.g.,][]{Werk2014,Fielding2017,Oppenheimer2018}.
The best estimates of the volume density of the
photoionized phase indicate that it is too rarified to be in pressure
equilibrium with a virialized hot gas halo 
\citep{Werk2014}.
Furthermore, it is predicted that such cool ``clumps'' will
be susceptible to Rayleigh-Taylor and Kelvin-Helmholtz instabilities
as they travel through the surrounding hot medium 
\citep{Schaye2007,HeitschPutman2009,Joung2012,Crighton2015,McCourt2015,Armillotta2016}.
In the absence of additional stabilizing mechanisms, hydrodynamical
simulations predict that these clumps are almost completely disrupted
on timescales similar to the cloud-crushing time:
\begin{equation}\label{eq.cct}
  t_{\rm crush} \sim \left ( \frac{\rho_{\rm clump}}{\rho_{\rm hot}} \right )^{1/2} \frac{R_{\rm clump}}{v_{\rm clump}}
\end{equation}
\citep{Jones1994,HeitschPutman2009,Joung2012,SchneiderRobertson2017},
with
($\rho_{\rm clump} /\rho_{\rm hot}$) the ratio of the densities of
the two phases, 
$v_{\rm clump}$ the relative velocity between the cool and hot media,
and 
$R_{\rm clump}$ the
size of the cool structure. Taken at face value, this relation
indicates that the survival time of the cool phase has a strong
dependence on both its size and kinematics. 

In spite of their apparent importance, however, empirical constraints on the sizes and
morphologies of the phases of the CGM are very few.  
Photoionization modeling of absorption systems detected along QSO sightlines with, e.g., the CLOUDY 
spectral synthesis code 
\citep{Ferland1998},
can in principle constrain the extent of the structure along the line
of sight.  However, such modeling requires that several ionic metal
transitions be observed, and the results are sensitive to the user's assumptions regarding the background ionizing radiation
field. Even in analyses of some of the richest absorption-line
datasets with coverage of numerous ionic species, the uncertainty in 
such size estimates are typically at least an order of magnitude 
\citep{Werk2014}.  Alternatively, multiple images of gravitationally-lensed QSOs can
have transverse separations ranging from less than a kiloparsec to
${>}100$ kpc scales, and if aligned behind  foreground absorption
can provide unique comparisons between the velocity structure and
strength of the system over these scales
\citep[e.g.,][]{WeymannFoltz1983,Smette1995,Monier1998,Rauch2002,Ellison2004,Chen2014}.
However, such special sightlines are very rare on the sky,
particularly if one also demands the presence of a close transverse
foreground galaxy whose redshift is known {\it a priori} \citep{Chen2010,Chen2014,Zahedy2016}.

Spectroscopy of spatially-extended background sources, such as bright
background galaxies, can also constrain the sizes of
foreground absorbers, especially when analyzed in tandem with
complementary QSO absorption spectroscopy (e.g., probing similar
foreground systems).  Galaxies that are sufficiently bright to enable
high-S/N spectroscopic coverage of foreground metal-line transitions
in the near-UV are rare; however, dense galaxy redshift surveys
can facilitate the selection of statistical samples of such exceptional
objects.  When identified
close in projection to foreground galaxies with
known redshifts, near-UV spectroscopy of the background 
sightlines may be used to target the coherence scale of the cool,
photoionized phase of the CGM in metal-line absorption transitions
(e.g., \ion{Mg}{2} $\lambda \lambda 2796, 2803$).  Background-foreground galaxy
pair spectroscopy has indeed been presented in several previous studies
\citep[e.g.,][]{Adelberger2005,Barger2008,Steidel2010,Rubin2010,Bordoloi2011,Diamond-Stanic2016,CookeOMeara2015,Lee2016,
Lopez2018,Peroux2018}.
However, until recently, no study has achieved the S/N necessary to
assess absorption equivalent widths associated with the photoionized CGM in more than
one or two individual foreground galaxy halos.


In Paper I of this series
(\citealt{GPG1}; hereafter \citetalias{GPG1}), we presented
spectroscopy obtained with the 
Keck/Low-resolution Imaging Spectrometer (LRIS) and the Very Large
Telescope (VLT)/Focal Reducer/Low-dispersion Spectrograph 2 (FORS2) 
of 72 projected pairs of galaxies at $0.4 \lesssim z
\lesssim 1.0$ drawn from the PRIsm MUlti-object Survey (PRIMUS;
\citealt{Coil2011primus}; \citealt{Cool2013}).  
Each individual background (b/g) galaxy spectrum is sufficiently deep to be
sensitive to \ion{Mg}{2} $\lambda 2796$ absorption 
with equivalent width $W_{2796} \gtrsim 0.5$ \AA, and as such provides consequential
constraints on CGM absorption associated with the corresponding foreground (f/g) halo.  The pairs probe impact parameters
as large as $R_{\perp} < 150$ kpc, but over two-thirds of the sample
has $R_{\perp} < 50$ kpc, thus probing regions known to exhibit the
strongest \ion{Mg}{2} absorption in complementary QSO sightline
experiments \citep[e.g.,][]{Chen2010}.  The f/g galaxies in these
pairs have stellar masses in the range $10^9 ~M_{\odot}< M_* <
10^{11.2}~M_{\odot}$, and lie predominantly along the star-forming sequence.

We then used these data to examine the relation between $W_{2796}$ and
$\mrperp$, demonstrating a negative correlation between these
quantities within $\mrperp < 50$ kpc.  We
explored the median relation between $W_{2796}$ and the intrinsic properties
of the f/g hosts, finding that greater $W_{2796}$  arises around
galaxies with higher star formation rates (SFR) and/or $M_*$.
Finally, we compared these measurements with a sample of $W_{2796}$
values obtained from studies of projected QSO-galaxy pairs in the
literature, finding that the median $W_{2796}$ observed 
toward both b/g galaxies and b/g QSOs
at a given
impact parameter around f/g galaxies of similar $M_*$ are
statistically consistent.

In the present work, we turn our focus from the mean and median CGM
absorption strengths to an examination of the dispersion in the
$W_{2796}$ distributions observed toward b/g galaxies and QSOs, and to
a detailed comparison of these distributions.  As we derive, 
the dispersion in $W_{2796}$ as a function of $\mrperp$ and intrinsic
host galaxy properties is dependent on the size of 
the b/g beam relative to that of the f/g absorber, and may therefore
constrain the latter quantity.
We begin our comparison by making
the assumption that the \ion{Mg}{2}-absorbing CGM as probed by 
our b/g QSO sample is the universal, or ``fiducial'' CGM --
i.e., that we are observing the same median $W_{2796}$ profile as a function of $\mrperp$
and $M_*$ toward both these QSOs and toward our PRIMUS b/g galaxies.
 In \S\ref{sec.fiducial} below, we develop a simple model for this fiducial
 $W_{2796}$ profile.  Then, in \S\ref{sec.biggerbeams}, we explore the
 relationship between the measured dispersion in this profile and the
 size of a given b/g beam relative to
 the sizes of the f/g absorbers in this fiducial CGM.  
Readers interested in the constraint on absorber size implied by the
 level of dispersion in $W_{2796}$  measured toward our b/g galaxy
sample may wish to focus on \S\ref{sec.result_coherence}. 
In \S\ref{sec.discuss} we 
discuss the
implications of these results for the 
physical nature of the 
\ion{Mg}{2}-absorbing CGM (\S\ref{sec.interp}), describe complementary constraints on its
small-scale structure (\S\ref{subsec.complementary_constraints}),
address the limitations of our analysis (\S\ref{subsec.assumptions}),
and 
discuss the lifetime and fate of this cool, photoionized
material (\S\ref{subsec.thermo}).  We offer concluding remarks in \S\ref{sec.conclude}.
We adopt a $\Lambda$CDM cosmology with $H_0 = 70~\rm
km~s^{-1}~Mpc^{-1}$, $\Omega_{\rm M} = 0.3$, and  $\Omega_{\Lambda} =
0.7$.  

\section{The Coherence Scale of $W_{2796}$}\label{sec.coherence}

Our goal is to perform a quantitative comparison of the dispersion in
$W_{2796}$ measurements obtained toward b/g QSOs and b/g galaxies to
constrain the physical scale of the f/g absorption.  
Our approach rests on a key assumption: that
$W_{2796}$ profiles obtained by assembling large samples of projected QSO-galaxy (or galaxy-galaxy)
pairs are representative of the ``fiducial'' CGM.  That is, we assume
that the dispersion in these $W_{2796}$ measurements is driven by the
spatial fluctuations in $W_{2796}$ in this fiducial CGM, rather than global
variations in CGM properties from one host galaxy to another (at a
given $M_*$, SFR, $\mrperp$, etc.).  This assumption has not yet been
justified empirically; moreover, the recent findings of \citet{Lopez2018}
suggest that this may overestimate the intrinsic $W_{2796}$ dispersion slightly.
Larger samples of galaxies with
extended or multiple b/g sightlines 
 (e.g., gravitationally lensed QSOs or galaxies;
\citealt{Chen2014,Zahedy2016,Lopez2018}) are needed to validate this picture.

Given this starting point, it follows that (as mentioned in
Section~\ref{sec.intro}) the \ion{Mg}{2}-absorbing CGM probed by our
PRIMUS b/g galaxy sample is the same, ``fiducial'' CGM probed by
existing projected QSO-galaxy pair samples at a similar epoch.
Analysis presented in 
\citetalias{GPG1} demonstrated that the $W_{2796}$ observed toward b/g
galaxies is larger around f/g hosts with higher SFR and/or $M_*$ at a given
$\mrperp$, in qualitative agreement with the results of projected
QSO-galaxy pair studies.  In addition, we demonstrated that the median
$W_{2796}$ observed toward our b/g galaxy sample is consistent with
that observed along b/g QSOs probing f/g halos over the same range in
stellar mass ($9.1 < \log M_*/M_{\odot} < 10.7$).  Although these
findings do not test the validity of our key assumption,
they are at least compatible with the concept of a fiducial CGM.
Moreover, this dependency of $W_{2796}$ on intrinsic galaxy properties
(as well as on $\mrperp$; \citealt{Chen2010,Nielsen2013}; \citetalias{GPG1})
implies that when comparing the dispersion in $W_{2796}$ among various
projected pair samples, we must account for 
 the differing $\mrperp$ distributions
and f/g galaxy properties of each dataset.

\subsection{A Fiducial Model for the Cool CGM}\label{sec.fiducial}

To facilitate this accounting, we start by developing a model for the
relationship between $\log W_{2796}$, $\mrperp$, and $M_*$ of the f/g
host.  This parametrization was 
first explored  in \citet{Chen2010b}, who
found that the inclusion of $M_*$ as an independent model variable significantly reduced the intrinsic scatter in the
relation between 
$\log W_{2796}$ and $\log \mrperp$ 
among their sample of 71 $W_{2796}$ measurements obtained from b/g QSO
spectroscopy probing ``isolated'' f/g galaxy halos at $z\sim0.25$ (over
$9~\mathrm{kpc} \lesssim \mrperp \lesssim 170~\mathrm{kpc}$).  
The demonstration of 
 a positive relation between $W_{2796}$ and $M_*$ among
the samples discussed in \citetalias{GPG1} suggests that this type of
model  may have a
similar effect in the present context.  
Given that \citetalias{GPG1} also presented evidence
for larger $W_{2796}$ around f/g hosts with larger SFR, a version
of the model  including a dependence on SFR (rather than $M_*$) might
similarly reduce the intrinsic scatter in the $\log
W_{2796}$ -- $\mrperp$  relation.  However,
because \citetalias{GPG1} did not identify a significantly stronger relationship
between $W_{2796}$ and one of these intrinsic host properties relative to
the other, for simplicity we choose to focus here on the potential
dependence on $M_*$ only.  Larger datasets
sampling the CGM of many more f/g hosts are needed to isolate the
relationships between these two correlated quantities and $W_{2796}$.
Recent studies have also suggested that $W_{2796}$ may depend on
the  azimuthal angle of the b/g
sightline relative to the f/g galaxy
\citep{Bordoloi2011,Bouche2012,Kacprzak2012}, implying an additional
reduction of the
true intrinsic scatter in the $W_{2796}$ distribution.  We test for
evidence of this dependence in our QSO-galaxy pair sample in 
Appendix~\ref{sec.azang}, again concluding that a larger
comparison dataset is required before such a dependence can be
productively incorporated into such a fiducial CGM model.

Our model thus simply includes a linear dependence on both $\mrperp$ and
$\log M_*$ as follows:
\begin{equation}
  \log \overbar{W}_{2796} = b + m_1 \mrperp + m_2 (\log M_*/M_{\odot}- 10.3),
\end{equation}
with $\log \overbar{W}_{2796}$ representing the predicted absorption strength, and
with the arbitrary offset of $10.3$ chosen to be close to the median
$M_*$ of the relevant datasets (described in more detail below).
As in \citet{Chen2010}, we adopt the  likelihood function
\begin{multline*}
  \mathcal{L}(\overbar{W}) = \left (\prod_{i=1}^{n} \frac{1}{\sqrt{2\pi
      s_i^2}} \exp \left
      \{-\frac{1}{2} \left [\frac{W_i - \overbar{W}}{s_i} \right
      ]^2 \right \} \right )\\
 \times \left (\prod_{i=1}^{m}
    \int_{-\infty}^{W_i} \frac{dW'}{\sqrt{2\pi
      s_i^2}} \exp \left
      \{-\frac{1}{2} \left [\frac{W' - \overbar{W}}{s_i} \right
      ]^2 \right \} \right ),  
\label{eq.likelihood}
\end{multline*}
here with
each value of $\overbar{W}$ equal to the value of $\log
\overbar{W}_{2796}$ given by the model at each f/g galaxy's $R_{\perp, i}$ and
$\log M_{*, i}$.  
As described in \citetalias{GPG1}, the first product includes direct
$\log W_{2796}$ measurements, and the second includes upper limits.
The Gaussian variance $s_i^2 = \sigma_i^2 + \sigma_C^2$, with
$\sigma_i$ representing measurement uncertainty and $\sigma_C$
representing the intrinsic scatter in the relation.

\begin{figure}
\begin{center}
\includegraphics[angle=0,width=\columnwidth,trim=20 20 20 -10,clip=]{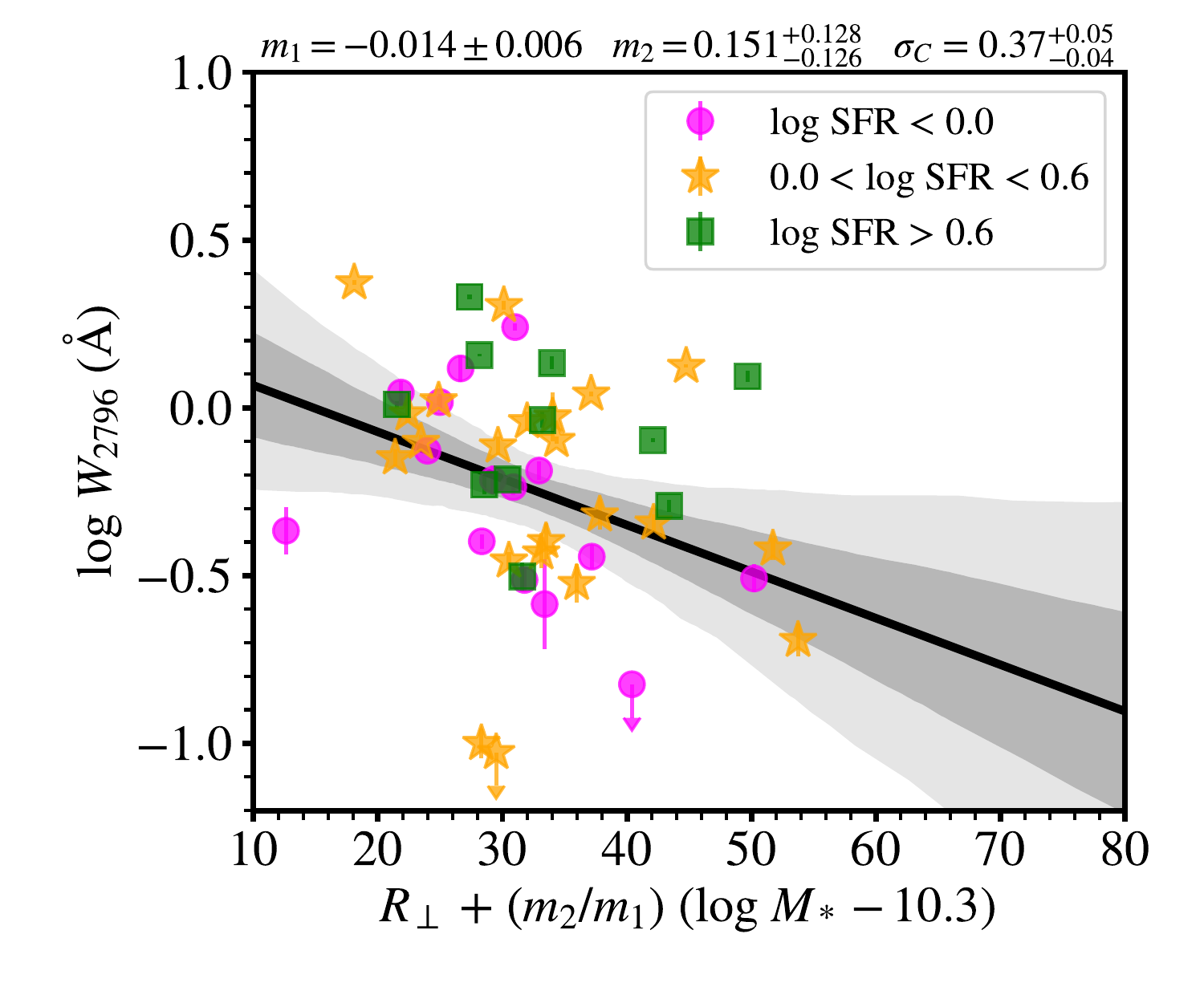}
\caption{$\log W_{2796}$ vs.\ a linear combination of $\mrperp$ and $\log
  M_*$ ($R_{\perp}^{\rm corr} = \mrperp + (m_2/m_1)(\log M_* -
10.3)$) for the QSO-galaxy pair sample we use
  to define the parameters of our fiducial model for the \ion{Mg}{2}-absorbing CGM.
  This includes pairs with f/g galaxies having $\mrperp < 50$ kpc and 
  that lie on the star-forming sequence as defined by \citet{Berti2016}.
  The median redshift of these systems is $z = 0.25$.
The best-fit values and $\pm34$th percentile
  probability intervals for relevant model parameters are listed above
  the top x-axis.  The quantity $R_{\perp}^{\rm corr}$ is calculated
  assuming $m_1$ and $m_2$ are equal to their best-fit values.
  Foreground galaxies having low, intermediate, and high values of SFR
  are indicated with magenta filled circles, orange stars, and green
  squares, respectively.  The black line shows the best-fit linear
  relation, and the dark and light gray contours indicate the inner
  $\pm34\%$ and $\pm47.5\%$ of the locus of curves
 obtained from random draws from the PPDF of the model.  
\label{fig.primus_qsofid}}
\end{center}
\end{figure}

We constrain the four free parameters of this model ($m_1$, $m_2$,
$b$, and $\sigma_C$) by fitting a subset of the QSO-galaxy pair
samples available in the literature chosen to define the
``fiducial'' CGM at this epoch.  These include the same datasets adopted
in \citetalias{GPG1}, namely those of \citet{Chen2010,Chen2010b} and
\citet{Werk2013}.  Both of these studies build their samples using f/g galaxies
whose redshifts are known {\it a priori}.  We cull these samples to
include only QSO sightlines which pass within $\mrperp < 50$ kpc of 
the corresponding f/g host
for consistency
with our primary PRIMUS pair sample.  
In addition, because some evidence suggests that the $W_{2796}$
profile of star-forming host halos differs from that of halos hosting
early-type galaxies \citep[e.g.,][]{Bordoloi2011}, and because the f/g
systems in our  PRIMUS pair sample are mostly star-forming,
we exclude any pairs with galaxies 
lying below the star-forming sequence as defined by \citet[][]{Berti2016}
(see Equation 1 in \citetalias{GPG1}).  The final fiducial QSO-galaxy pair
sample 
is shown in Figure~\ref{fig.primus_qsofid}, and in total includes 39
measurements from \citet{Chen2010b} and 11 measurements from
\citet{Werk2013}.  

Following the methodology of \citetalias{GPG1}, we use the Markov
Chain Monte Carlo technique to sample the posterior probability
density function (PPDF) for our model given these data (using the software
package \texttt{emcee}; \citealt{DFM2013}).  We use uniform
probability density priors over the ranges $-5.0 < m_1 < 5.0$, $-5.0 <
m_2 < 5.0$, $-10.0 < b < 10.0$, and $-10.0 < \ln \sigma_C < 10.0$.  As
in \citetalias{GPG1}, we  find that chains generated by 100
``walkers'' each taking 5000 steps thoroughly sample the
PPDF.  We again adopt the median and $\pm34$th-percentiles of the
marginalized PPDFs as the best-fit value and uncertainty for each parameter.



Figure~\ref{fig.primus_qsofid} shows the outcome of this procedure.
The best-fit parameter values with uncertainties are listed at the top
of the plot.  The x-axis shows a linear combination of the two
independent variables, $R_{\perp}^{\rm corr} = \mrperp + (m_2/m_1)(\log M_* -
10.3)$, with 
$m_1$ and $m_2$ set at their best-fit values.
The thick black line shows the best-fit relation.  The dark and
light gray contours show the $\pm1\sigma$ and $2\sigma$ uncertainty ranges in $\log
\overbar{W}_{2796}$, 
estimated by drawing 1000 random sets of parameters from the PPDF,
calculating $\overbar{W}$ for each set, and then determining the
inner $\pm34$th and $\pm47.5$th percentile values of $\overbar{W}$ as
a function of $R_{\perp}^{\rm corr}$.

The fitted slope $m_2$ 
is suggestive of a marginally statistically-significant dependence on
$\log M_*$, with $m_2 = 0.151^{+0.128}_{-0.126}$ larger than zero at a
level of ${\sim} 1.2\sigma$.  
The best-fit value and uncertainty interval for $\sigma_C$, on the
other hand, is indicative of a high level of intrinsic scatter around
the best-fit  linear relation.
The data points are color-coded by SFR as indicated in the legend.
The distributions of $\log
W_{2796}$ values with respect to the best-fit relation are broadly
consistent 
among these subsamples, suggesting that if there is
an additional dependence of $W_{2796}$ on SFR, this sample will not
usefully constrain it. 

\begin{figure}
\begin{center}
\includegraphics[angle=0,width=3.2in,trim=20 20 20
20,clip=]{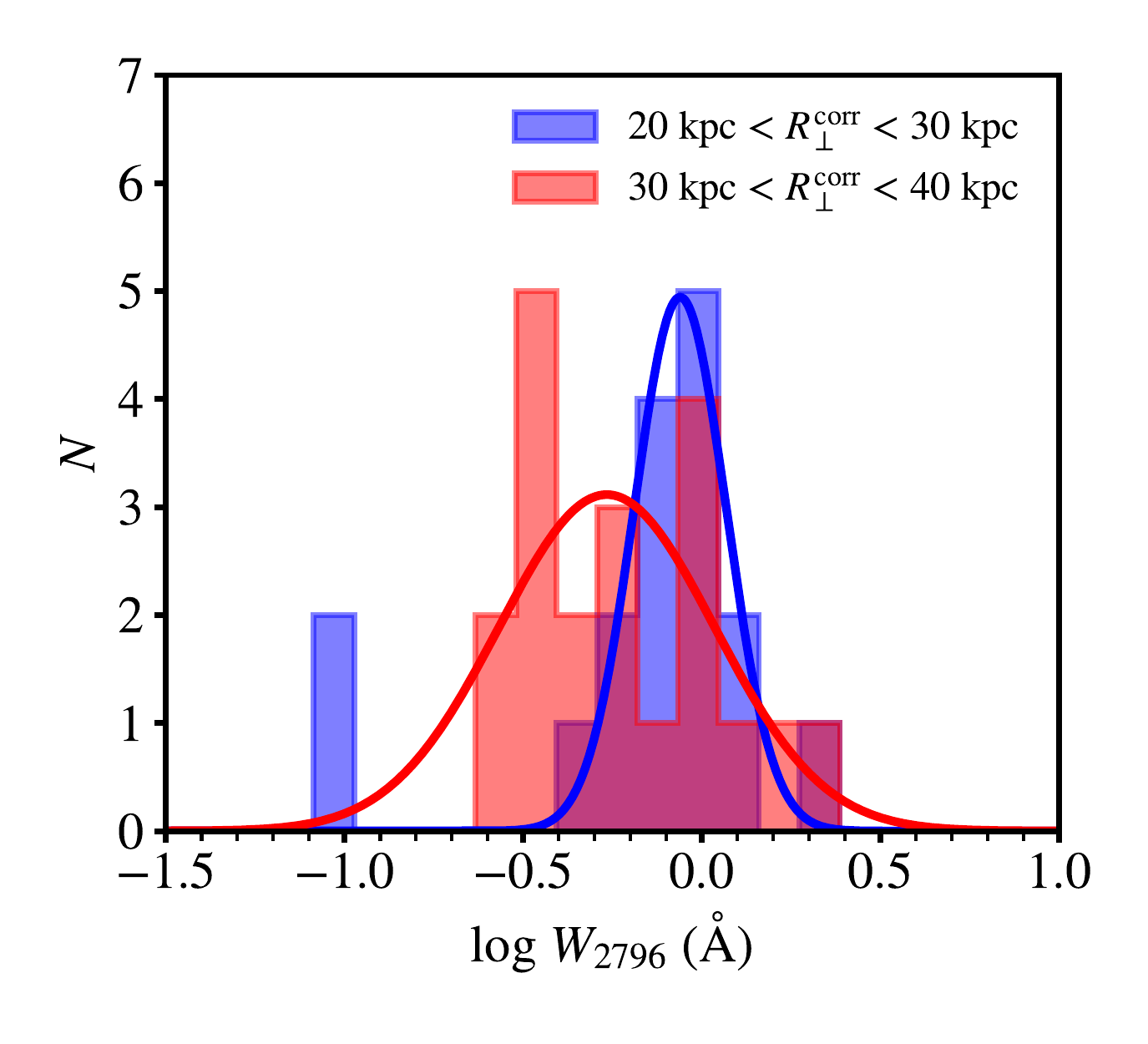}
\caption{The distribution of $\log W_{2796}$ values in our
  QSO-galaxy comparison sample in two bins of the quantity
  $R_{\perp}^{\rm corr}$, with $R_{\perp}^{\rm corr} = \mrperp + (m_2/m_1)(\log M_* -
10.3)$, and with $m_2/m_1 =  -10.8$.  Two of these constraints on
$\log W_{2796}$ are upper limits, and are included in the histograms
at the values of these limits.
The blue and red curves are
fits of Gaussian functions to the $\log W_{2796}$
distributions at 20 kpc $<
R_{\perp}^{\rm corr} <$ 30 kpc and 30 kpc $< R_{\perp}^{\rm corr} <$ 40 kpc,
respectively.  The Gaussians are shown only to demonstrate the form of
the scatter in $\log W_{2796}$ we assume. Their fitted parameters 
are not used in our
analysis.
\label{fig.primus_ewdisp}}
\end{center}
\end{figure}

We use this model and these best-fit parameters to define the 
form of and intrinsic scatter in $\log \overbar{W}_{2796}$ as a function of
$\mrperp$ and $\log M_*$; i.e., our 
 ``fiducial'' \ion{Mg}{2}-absorbing CGM model.  Our
 constraints on all of these parameters (including the intrinsic scatter,
 $\sigma_C$) will
 be leveraged in the analysis to follow.  Here we remind the reader
 that we have assumed a Gaussian form for the variance in $\log
 \overbar{W}_{2796}$, implying that the probability distribution of a
 measurement $\log
 W_{2796, i}$ at particular values of $R_{\perp, i}$ and
 $\log M_{*, i}$ is also a Gaussian centered at $\log
 \overbar{W}_{2796, i}$ \citep{Hogg2010}.  This assumption of
 Gaussianity in $\log W_{2796}$ (and hence, lognormality in
 $W_{2796}$) has not yet been justified; however, we persist in this
assumption for the following reasons.  
First, we lack the measurements needed to empirically constrain the
form of the $W_{2796}$ distribution, and therefore consider the choice
between normal and lognormal distributions arbitrary.  
Second, while
negative values of $W_{2796}$ may be measured in instances of noisy
spectroscopy, the true $W_{2796}$ due to diffuse \ion{Mg}{2} ions foreground
to a bright background source will always be non-negative.  
A lognormal distribution is consistent with this constraint.
A test of this assumption may in principle be performed 
as demonstrated in Figure~\ref{fig.primus_ewdisp}: here, we select 
subsamples of the QSO-galaxy comparison dataset in two narrow ranges
in the quantity $R_{\perp}^{\rm corr} = \mrperp + (m_2/m_1)(\log M_* -
10.3)$ and show the distribution of $\log W_{2796}$ in each.  
We cannot use these subsamples to quantitatively disfavor a lognormal
relative to a Gaussian model (or vice versa) for the $W_{2796}$ distribution due
to the
small number of measurements.  We encourage future studies
with larger samples to address this issue.



\subsection{Observing the Fiducial CGM Toward Larger Background Beams}\label{sec.biggerbeams}

The PPDF for the parameters of the fiducial model
developed in the previous subsection may now be used to generate  
new samples of $W_{2796}$ measurements that would be observed along b/g
QSO sightlines given any arbitary set of f/g
galaxy $M_*$ and impact parameter values.
Our ultimate goal is to constrain the sizes of \ion{Mg}{2} absorbers
by comparing this model to our measurements of $W_{2796}$
toward b/g galaxies (i.e., sources that emit UV continuum light over significantly
larger areas than QSOs).  We begin this comparison by considering how
the distribution of $W_{2796}$ values 
would change if such a fiducial CGM, once generated by this model,
could be {\it reobserved} using b/g
light sources of increasing size.

Perhaps the simplest configuration to consider is that of an extended background
source consisting of numerous point-like (i.e., QSO-like) sources of equal
intensity covering a projected area $A_{\rm G}$.  Picture the light from  these point sources passing
through CGM material and  being absorbed by
foreground \ion{Mg}{2} ions.  The strength of this absorption need not
be the same along all of these sightlines; here, we assume the
absorption observed toward each sightline $j$ yields an equivalent
width $W_{2796}^j$.  
In this case, the
equivalent width observed along a spectrum which integrates the light from
all of these point sources, $W_{2796}(A_{\rm G})$, is equal to the
average of all $W_{2796}^j$, $W_{2796}(A_{\rm G}) = \frac{1}{N} \sum_{j=1}^{N}  W_{2796}^j$.
This holds for 
any number of \ion{Mg}{2}-absorbing structures with any 
column density and velocity distributions along these lines of sight, so
long as the structures give rise to the total specified equivalent
widths ($W_{2796}^1$, $W_{2796}^2$, etc.).\footnote{If the point
  sources instead emit continua with varying intensities, the final
  observed $W_{2796}(A_{\rm G})$ will be a continuum flux-weighted
  average of the equivalent widths along each sightline.}

\begin{figure}
\begin{center}
\includegraphics[angle=0,width=3.2in,trim=20 20 20
15,clip=]{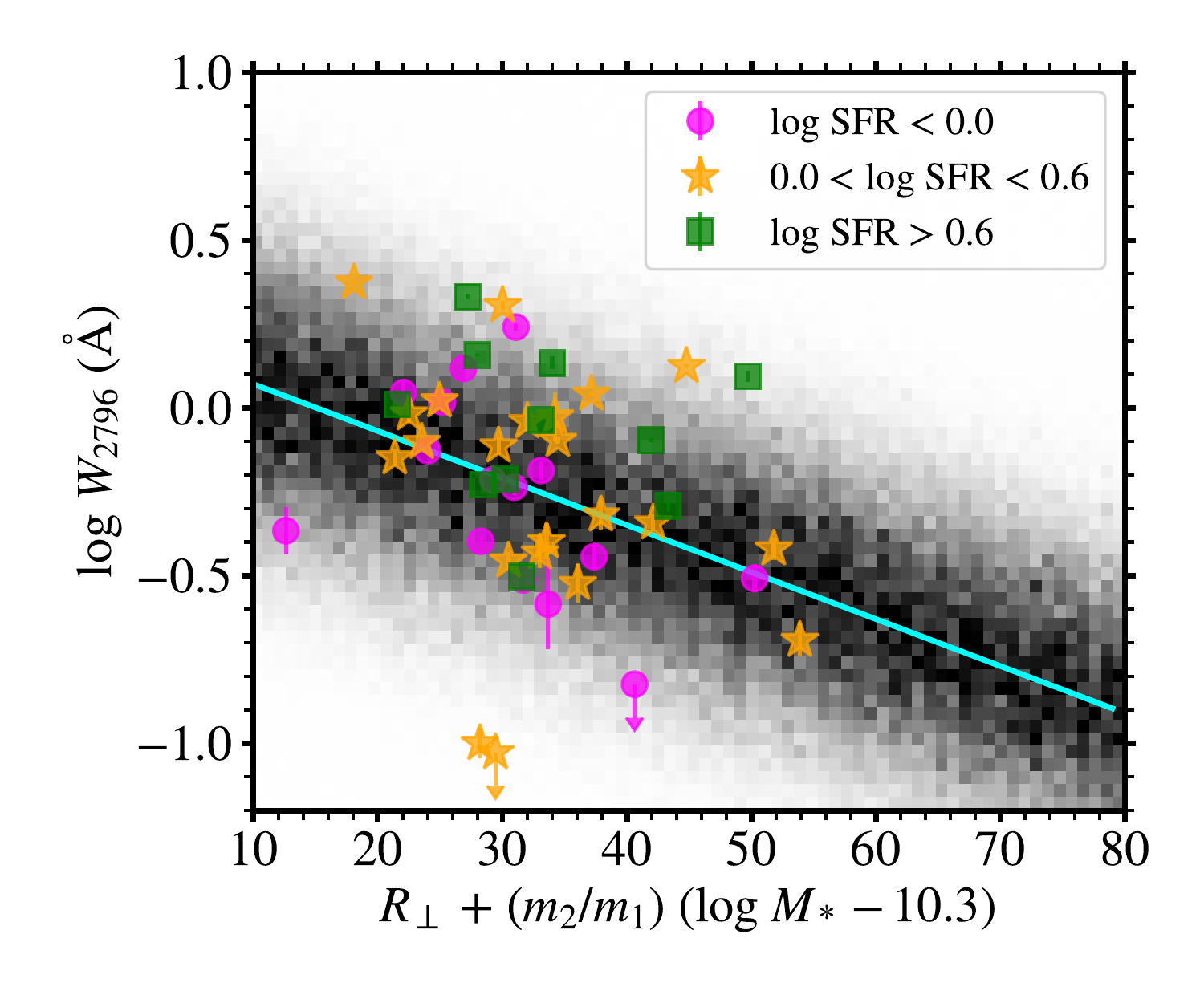}
\caption{
$\log W_{2796}$ vs.\ $R_{\perp}^{\rm corr}= \mrperp + (m_1/m_2)(\log M_* - 10.3)$ for our simulated
observations of the fiducial \ion{Mg}{2}-absorbing CGM.
$R_{\perp}^{\rm corr}$ 
is calculated as described in the Figure~\ref{fig.primus_qsofid}
caption.  The two-dimensional histogram indicates the number of simulated measurements
in each bin, normalized by the maximum number per bin within each 
column.  The grayscale varies linearly with point density, with the
darkest shading indicating the bins with the highest density.   The
cyan line shows the best-fit fiducial model as determined in
\S\ref{sec.fiducial}, and the large points show the subset of the
QSO-galaxy pair sample used to constrain this model.  The
points are colored as described in the legend and the
Figure~\ref{fig.primus_qsofid} caption.
\label{fig.primus_fidgenerate}}
\end{center}
\end{figure}

To make use of these mock extended b/g sources, we now refer to our
fiducial model to generate a large sample of $W_{2796}$ measurements (the
distribution of which, by definition, is consistent with the
$W_{2796}$ distribution of our QSO-galaxy pair sample).
We assume that the true parameters defining the relationship between
$\mrperp$, $\log M_*$, and $\log W_{2796}$ are equal to the best-fit
values of $m_1$, $m_2$, $b$ and $\sigma_C$ as determined in
\S\ref{sec.fiducial}.  We then 
evenly sample the range in $R_{\perp}^{\rm corr}$ occupied by the
QSO-galaxy pair dataset.  For a given $R_{\perp}^{\rm corr}$ value, 
we calculate the corresponding $\log \overbar{W}_{2796}$
implied by the fiducial model, and then generate 5000 random draws from a
standard normal distribution centered at $\log \overbar{W}_{2796}$
with a standard deviation equal to $\sigma_C$.  We
repeat this process at every $R_{\perp}^{\rm corr}$, and show the
resulting distribution of $\log W_{2796}$ measurements 
in Figure~\ref{fig.primus_fidgenerate}.
Here, the two-dimensional histogram indicates the number of realized $\log
W_{2796}$ values in each bin, normalized by the maximum number per bin
in the corresponding histogram column.  
The grayscale varies linearly with density, with the darkest shading
indicating the most frequently sampled bins.
We note that formally, the variance in this simulated dataset ($\sigma_C^2$) is slightly lower
than in the observed dataset ($s_i^2 = \sigma_i^2 + \sigma_C^2$), as
a proper realization of the observations would include an estimate of
the typical measurement uncertainty ($\sigma_W$) and draw from a
Gaussian with a dispersion $\sqrt{\sigma_W^2 + \sigma_C^2}$.  
However, the median value of $\sigma_i$ in this sample is
${\sim}0.025$, over an order of magnitude lower than $\sigma_C$, such
that its inclusion in the variance would yield a negligible difference in the final
realized sample.  Figure~\ref{fig.primus_fidgenerate} shows the
QSO-galaxy pair sample with colored points as in 
Figure~\ref{fig.primus_qsofid} for comparison to our simulated
dataset.  The best-fit fiducial relation is shown in cyan.


We may now use this simulated dataset to predict the distribution of a
{\it new} set of observations made with our simple extended background
beam.  First, we must consider a new quantity: the projected area of
each \ion{Mg}{2} absorber, $A_{\rm A}$.
Here, we use  ``absorber'' or ``structure'' to describe a system that gives
rise to the same $W_{2796}$ across its entire projected area.  
By definition, then, the edges of
every absorber just touch those of the neighboring
structures (so that there is no ``empty space'' nor any overlap between absorbers).
We remind the reader that the \ion{Mg}{2} absorbers we are considering
are strongly saturated, such that $W_{2796}$ is likely more closely
correlated with the velocity dispersion and number of absorbing
clouds along the sightline than the \ion{Mg}{2} column density.
We also note that although many of the absorbers in our sample
  have a 1:1 \ion{Mg}{2} doublet ratio without exhibiting line-black troughs 
 (see Table 4 and Figure 18 in \citetalias{GPG1}), we
  are not able to usefully constrain the gas covering fraction via absorption
  line analysis due to the low spectral resolution of the data.

 The standard way of describing the $W_{2796}(A_{\rm G})$ that
  would be observed for {\it one}
  of these absorbers (that is, assuming all neighboring absorbers have
  ${W}_{2796}^{k} = 0$ \AA)
is to write

\begin{equation}
  {W}_{2796}(A_{\rm G}) = C_f \int (1 - e^{-\tau_{\lambda}}) d\lambda, 
\end{equation}

\noindent where $\tau_{\lambda}$ is the optical depth in the
\ion{Mg}{2} 2796 transition, and the integral above is the equivalent
width that would be observed if the b/g source were fully covered
(which we are calling here  ${W}_{2796}^{k}$).  In the case that
the velocity structure and column density is coherent over $A_{\rm A}$
and $A_{\rm A} \le A_{\rm G}$, then
the covering fraction $C_f = A_{\rm A} / A_{\rm G}$, and $W_{2796}
(A_{\rm G}) = C_f
{W}_{2796}^{k}$.
Moreover, an alternative scenario in which the column density and velocity structure of the
absorption vary significantly within $A_{\rm A}$ (in such a way
that they produce the same ${W}_{2796}$ at every location within
$A_{\rm A}$) is also
permitted in this framework, and would yield equivalent results.


For our fiducial model, we assume that all absorbers have the same
area ($A_{\rm A}$), and we model the transverse size of each absorber using a square
geometry.  This allows us to populate a continuous two-dimensional
cross section through the CGM.
While there exist few empirical constraints on the shape and
size of even individual \ion{Mg}{2} absorbers, much less on these
quantities as a function of $W_{2796}$,
we expect that these
 two assumptions are significant oversimplifications of the
physical layout of cool CGM material.  We adopt them here to enable our
demonstration of the qualitative effect of using extended background beams
on the observed $W_{2796}$ distribution.  
We will discuss requirements
for increasing the realism of this model in \S\ref{subsec.assumptions}.

\begin{figure}
\begin{center}
\includegraphics[angle=0,width=\columnwidth,trim=20 20 20
20,clip=]{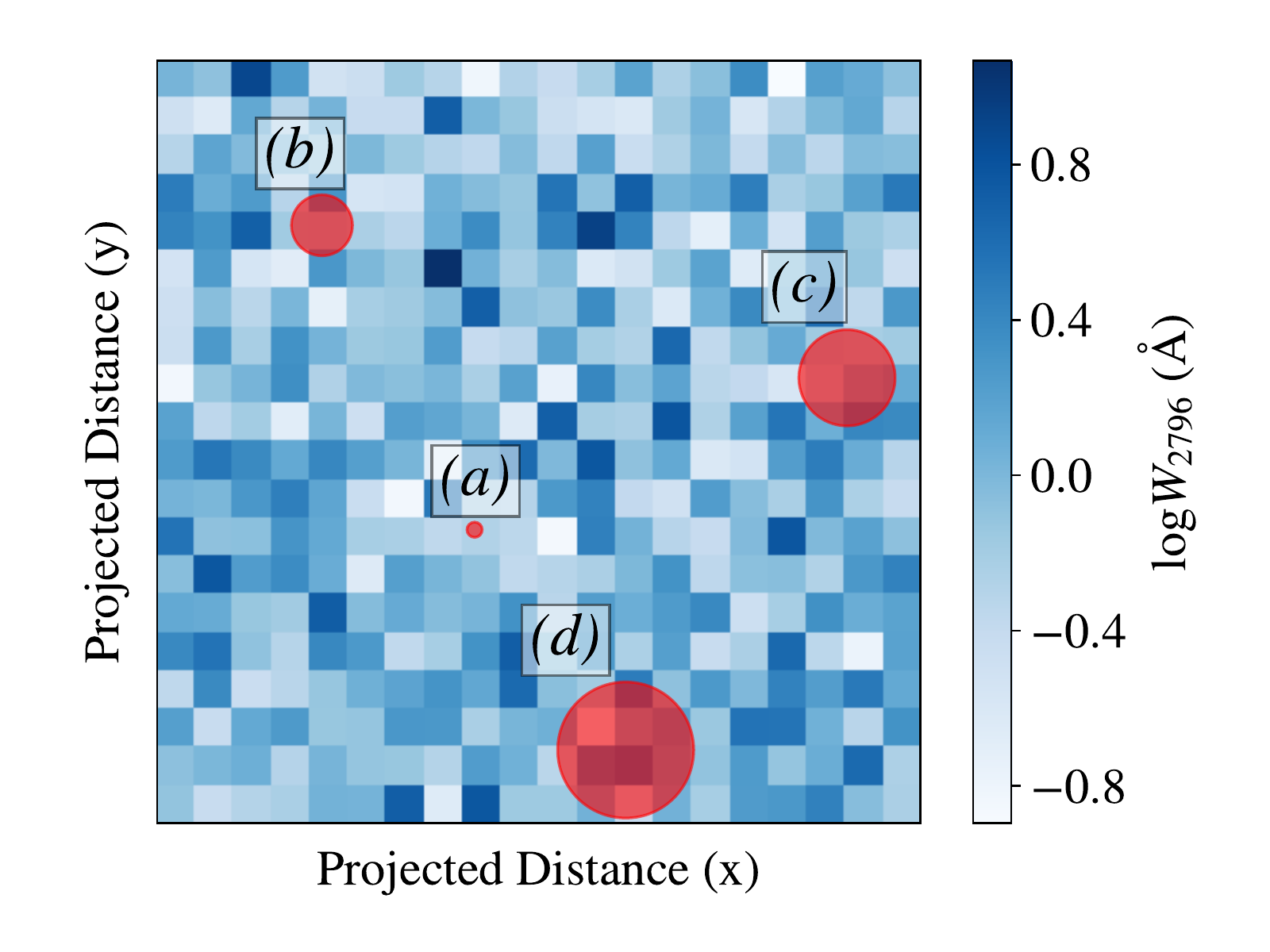}
\caption{Realization of the $W_{2796}$ distribution within a small
  region of our model CGM.  Each pixel in the map represents a different
  absorber and is color-coded by its $W_{2796}$ as indicated in the
  color bar.  This particular realization is generated from a Gaussian
  $\log W_{2796}$ distribution centered at $\log \overbar{W}_{2796} = 0$,
  and is thus meant to be representative of the CGM at $\mrperp^{\rm corr}
  \sim 15$ kpc.  The red circles represent extended b/g sources of
  various sizes, and their labels indicate the corresponding
  two-dimensional $W_{2796}$ distributions in the four panels of
  Figure~\ref{fig.primus_galfid}.  Note that the choice of physical size of each
  absorber (or pixel) in this toy model is unimportant; it is the
  ratio of the projected area of the b/g beam to the area covered by each
  absorber ($x_{\rm A}$) that affects the observed $W_{2796}$ distribution.
\label{fig.primus_2dcgm}}
\end{center}
\end{figure}

We now consider the
spatial distribution of absorbers within a small region of a halo  (say a $10\times10$ kpc
projected ``patch'' at $\mrperp^{\rm corr}\sim15$ kpc).  We may populate
this region with absorbers by generating random draws from the standard
normal distribution of the fiducial CGM centered at the
$\log\overbar{W}_{2796}$ appropriate for this value of $\mrperp^{\rm
  corr}$.  We place these absorbers at random
locations within the region.  An example of such a realization is shown
in Figure~\ref{fig.primus_2dcgm}.  This $20\times20$ pixel box is
populated with absorbers with a Gaussian $\log W_{2796}$ distribution
having a mean $\log\overbar{W}_{2796} = 0$ and a standard deviation
$\sigma_C$.  We then measure the absorption observed toward circular
beams placed at random locations behind this region.  
 For the purposes of this illustration, each absorber is $0.5\times
0.5$ kpc in area, and we are assuming that the gradient in
$\log\overbar{W}_{2796}$ over the range in $\mrperp^{\rm corr}$ covered by
this patch is negligible.
In the case that the size of the b/g beam is much smaller than
the projected area of the f/g structures ($A_{\rm G} \ll A_{\rm A}$, or the
ratio $x_{\rm A} \equiv \frac{A_{\rm G}}{A_{\rm A}} \ll 1$), as
demonstrated with the beam marked {\it (a)} in the Figure,
the observed absorption $W_{2796} (A_{\rm G})$ will almost always equal
$W_{2796}^{k}$, the equivalent width of the particular structure
probed.  
There may be some instances in which the b/g beam is placed
behind the edges of two absorbers, such that $W_{2796} (A_{\rm G})$
is a weighted average of $W_{2796}^{k=1}$ and
$W_{2796}^{k=2}$.  However, this will be rare given our condition for
the relative sizes of the beams vs.\ the absorbers.  In this case, the
observed distribution of $\log W_{2796}$ will be similar to that
observed toward QSOs themselves; i.e., consistent with our realization
of the fiducial CGM generated above.

\begin{figure*}
\begin{center}
\includegraphics[angle=0,width=7in,trim=20 20 20
20,clip=]{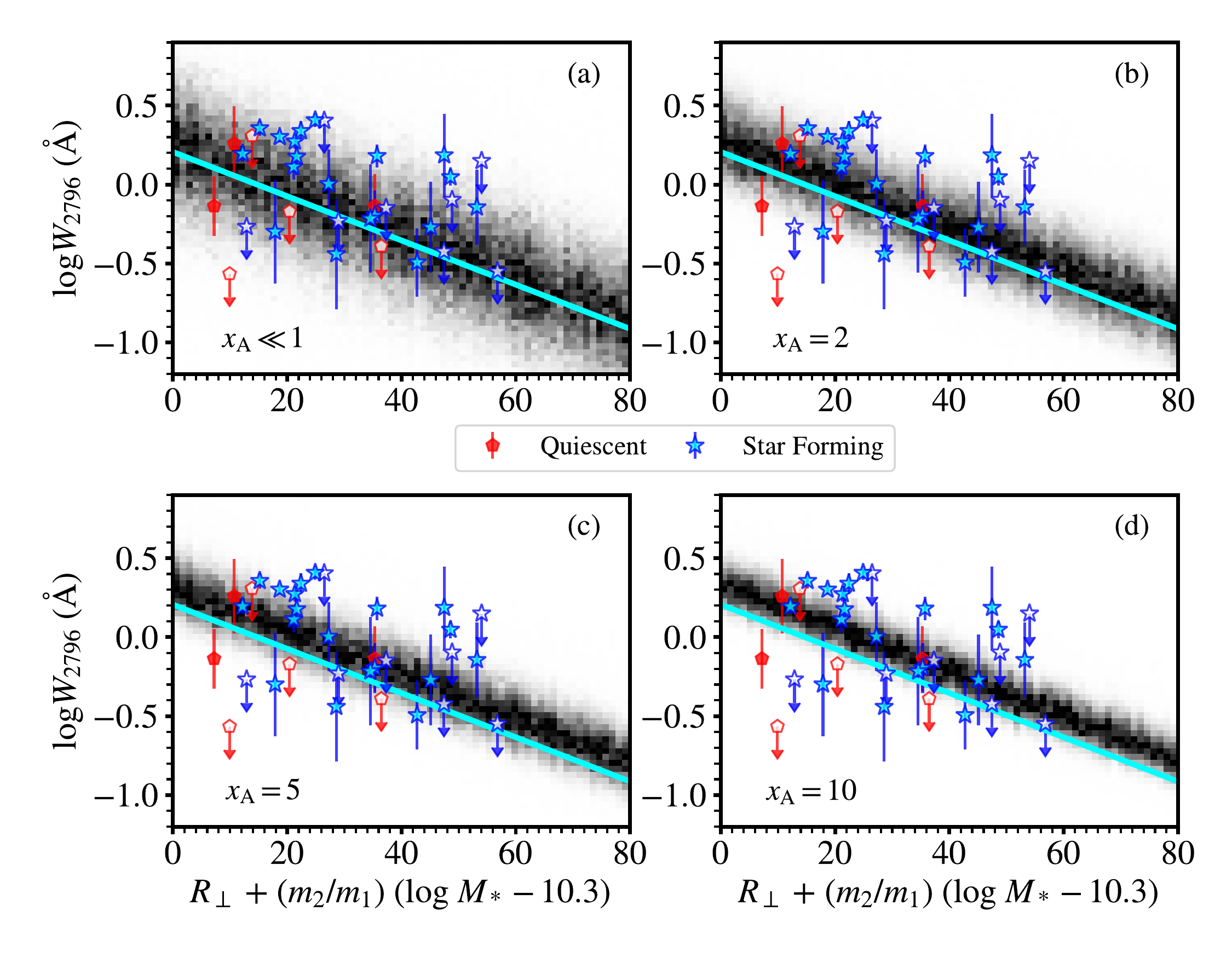}
\caption{The fiducial \ion{Mg}{2}-absorbing CGM model as observed
  toward extended b/g beams.  Each column of each grayscale histogram is calculated by
  generating a projected ``patch'' of randomly distributed \ion{Mg}{2}
  absorbers with area $A_{\rm A}$ selected from a Gaussian $\log W_{2796}$ distribution
  having a mean $\log \overbar{W}_{2796}$ corresponding to
  $\mrperp^{\rm corr}$ of that column (as demonstrated in Figure~\ref{fig.primus_2dcgm}).
  We then ``observe'' each patch toward numerous randomly-placed b/g
  beams with areas ($A_{\rm G}$) chosen as indicated with the ratio
  $x_{\rm A} \equiv \frac{A_{\rm G}}{A_{\rm A}}$
  at the lower left
  in each panel (and by the red circles in
  Figure~\ref{fig.primus_2dcgm}).  The histograms show the number of 
  simulated
  $\log W_{2796}$ measurements per bin, normalized as described in
  Figure~\ref{fig.primus_fidgenerate}. 
 The cyan line shows the best-fit
  fiducial linear model.  The colored
  points show $\log W_{2796}$ measured along PRIMUS b/g galaxy
  sightlines that lack bright AGN and with $R_{\perp} < 50$ kpc.  Pairs with star-forming f/g
  galaxies 
are indicated with cyan/blue stars, and 
pairs in
  which the f/g galaxy is quiescent are indicated with red pentagons.
 Pairs with star-forming and quiescent f/g hosts that yield upper limits on $\log W_{2796}$ are shown with open
  blue stars and open red pentagons, respectively.
  The dispersion in the $\log W_{2796}$ distribution is predicted to
  decrease as the ratio $x_{\rm A}$ increases.  The
  $\log W_{2796}$ distribution observed toward PRIMUS b/g galaxies appears
  consistent with that expected for a small value for this ratio
  ($x_{\rm A} \ll 1$), suggesting that the strength of
  \ion{Mg}{2} absorption varies on a scale larger than these b/g beams.
\label{fig.primus_galfid}}
\end{center}
\end{figure*}

This  is demonstrated in
Figure~\ref{fig.primus_galfid}a.  To generate the two-dimensional
histogram shown, we again evenly sample the range in
$\mrperp^{\rm corr}$ covered by the QSO-galaxy pair sample, 
and simulate a CGM patch at each $\mrperp^{\rm corr}$ as described above with the
appropriate $\log\overbar{W}_{2796}$ value.  To ensure a thorough
sampling of the Gaussian, each patch includes $50\times50$ absorbers
(or pixels).  
We adopt a beam radius of $r_{\rm G} = 0.2$ pixels, such that $A_{\rm
  G} = \pi r_{\rm G}^2 = 0.13 A_{\rm A}$ in this case. 
We then place 2500 of these beams in random locations ``behind'' every patch,
recording the $W_{2796}(A_{\rm G})$ observed toward each beam.  The
two-dimensional histogram in Figure~\ref{fig.primus_galfid}a shows the
distribution of these $W_{2796}(A_{\rm G})$ values for every patch,
normalized as in Figure~\ref{fig.primus_fidgenerate}.  
 Note that we generate a new patch for each step along the
  $x$-axis of size $\Delta \mrperp^{\rm corr}\approx 1$ kpc, rather than adopting
  the same $\log W_{2796}$ distribution over an $\Delta \mrperp^{\rm corr}$
  interval of several kpc as in the illustration shown in Figure~\ref{fig.primus_2dcgm}.
Our PRIMUS
galaxy pair $W_{2796}$ measurements, including only pairs having
$\mrperp < 50$ kpc, those in which the
b/g galaxy is not host to a bright AGN, and
absorption measurements that are unaffected by blending, are shown with large colored
points (\citetalias{GPG1}).  The best-fit relation is again shown with
a cyan line.  
As expected, the histogram  in
Figure~\ref{fig.primus_galfid}a is just slightly more narrowly
distributed about this best-fit model than the
histogram shown in 
Figure~\ref{fig.primus_fidgenerate}.

As the size of the b/g beam approaches the size of the f/g absorbers,
the frequency of sightlines probing more than one absorber increases.
If $A_{\rm G} > A_{\rm A}$, {\it every} b/g sightline will
intercept more than one absorber.  
To demonstrate this, we repeat the above exercise adopting b/g beams
such that $x_{\rm A} = 2$, $5$, and $10$ (as indicated by the
beams {\it(b)}, {\it(c)} and {\it(d)} in Figure~\ref{fig.primus_2dcgm}), and show the
resulting $\log W_{2796}$ distributions in
Figure~\ref{fig.primus_galfid} panels (b), (c), and (d),
respectively.  
 In effect, for each beam we are performing the sum
$W_{2796}(A_{\rm G}) = \sum_{k=1}^{N}  C_f^k W_{2796}^{k}$, where
the covering fraction is adjusted from its standard value of 
$C_f^k = A_{\rm A} / A_{\rm G}$ for absorbers 
with edges that overlap those of the beam. 
The scatter in these distributions is reduced as the
b/g beam increases in size; in addition, because large beams
are measuring the arithmetic mean of $W_{2796}$ values drawn from a
lognormal distribution, the $\log W_{2796}$ which occurs
with the highest frequency
at each $R_{\perp}^{\rm corr}$ lies slightly above the original fiducial
best-fit relation.  

 We may therefore constrain the ratio $x_{\rm A}$ by
comparing these simulated two-dimensional $\log W_{2796}$ distributions with
the dispersion in $\log W_{2796}$ measured in our PRIMUS galaxy pair
dataset. 
For consistency with the QSO-galaxy pair subsample used to
construct our fiducial model, we must limit our comparison to include
only the 27 galaxy pairs from \citetalias{GPG1} with f/g galaxies
that are star-forming (shown with cyan and blue symbols in
Figure~\ref{fig.primus_galfid}; 
we also show measurements for pairs with
passive f/g galaxies in red for completeness).  Examining the figure
by eye, it is clear that the galaxy pair sample $\log W_{2796}$
measurements exhibit a relatively
high level of dispersion about the best-fit fiducial model, apparently
comparable to that of the two-dimensional histogram in panel (a).

\begin{figure*}
\begin{center}
\includegraphics[angle=0,width=7in,trim=20 20 20
0,clip=]{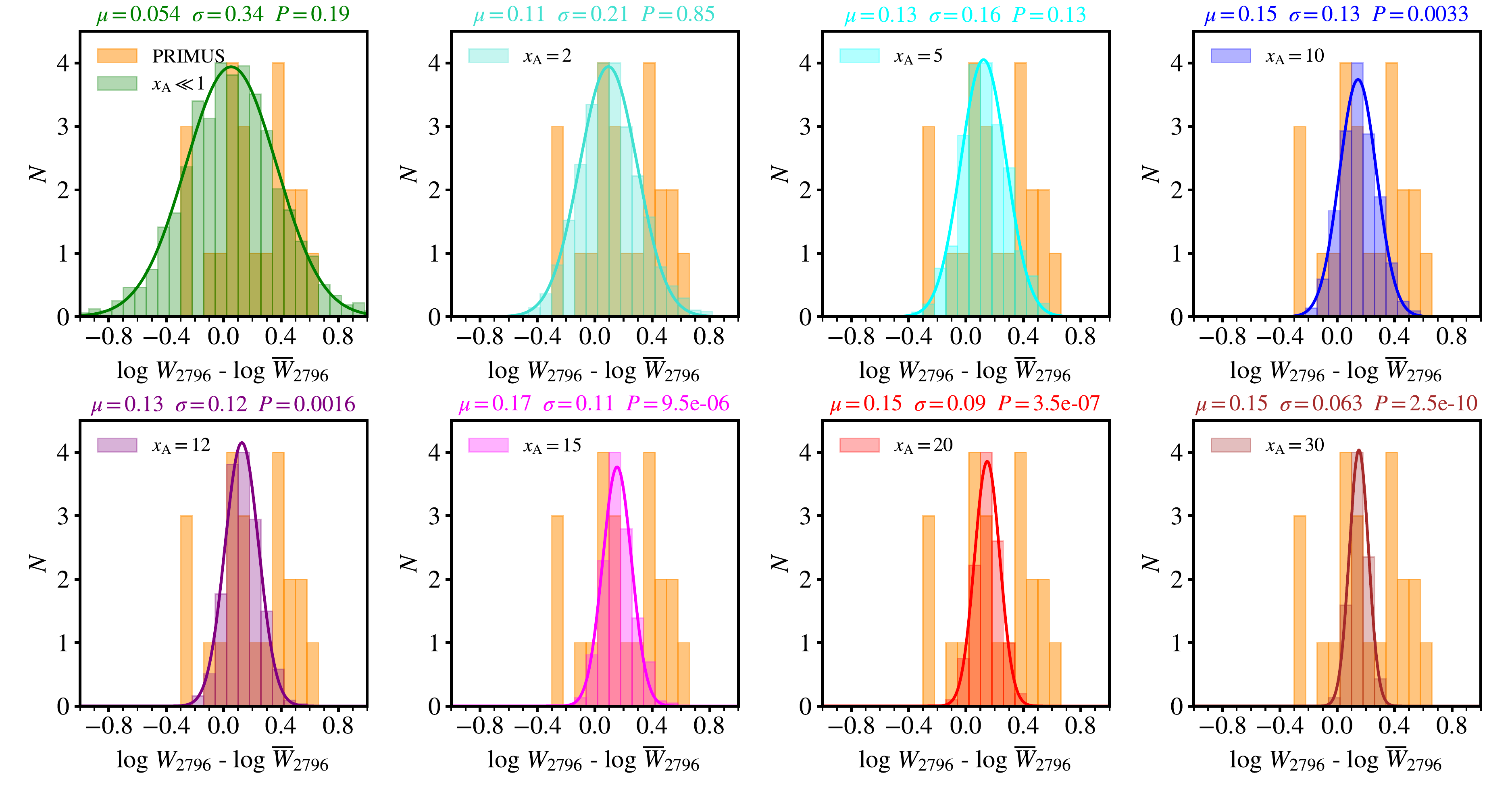}
\caption{Distribution of the offsets ($\Delta \log W_{2796}$) between the best-fit fiducial
  \ion{Mg}{2}-absorbing CGM relation 
  and our $\log W_{2796}$ measurements toward PRIMUS b/g galaxies probing
  the halos of star-forming f/g  galaxies (orange histograms).  In cases
  for which $\log W_{2796}$ is an upper limit, the offset is included
  in the plotted distribution only if it is $< 0.1$.  The green, turquoise,
  cyan, and blue histograms (top panels) show the  distribution of
  $\Delta \log W_{2796}$ offsets for each of the simulated datasets shown in
  Figure~\ref{fig.primus_galfid}; i.e., assuming a fiducial
  \ion{Mg}{2} CGM observed with b/g beams having  $x_{\rm A} \ll 1$, $x_{\rm A} = 2$, $x_{\rm A} = 5$, and
  $x_{\rm A} = 10$, respectively.  
The purple, magenta, red, and brown histograms (bottom panels) show
simulated distributions of $\Delta \log W_{2796}$  offsets
assuming $x_{\rm A} = 12$, $x_{\rm A} = 15$, $x_{\rm A} = 20$, and
$x_{\rm A} =30$
(not shown in Figure~\ref{fig.primus_galfid}).
The colored curves show
  Gaussian fits to each distribution.  The mean ($\mu$) and
  standard deviation ($\sigma$) of each simulated dataset is printed
  above the corresponding panel, along with 
 the $P$-value obtained from a log-rank test of the consistency
 between the observations and these simulated distributions. 
\label{fig.primus_1dewdist}}
\end{center}
\end{figure*}

To quantify this dispersion, we calculate the offset between the
best-fit fiducial relation and each $\log W_{2796}$ measurement ($\Delta \log
W_{2796} = \log W_{2796} - \log \overbar{W}_{2796}$) and show the
distribution of these values with the orange histograms in
Figure~\ref{fig.primus_1dewdist}.  Only measurements of the CGM around
star-forming hosts
are included here; in addition, in cases for which
$W_{2796}$ is an upper limit, the offset of this limit is shown only
if $\Delta \log W_{2796} < 0.1$.  We calculate the same offsets for
the simulated CGM datasets shown in Figure~\ref{fig.primus_galfid},
with the green, turquoise, cyan, and blue histograms showing the
dispersion in datasets predicted for b/g beams with 
$x_{\rm A} \ll 1$, $x_{\rm A} =2, 5$ and 10, respectively.
Here, we  simulate additional datasets that assume
b/g beams having 
$x_{\rm A} = 12, 15, 20$ and 30, and show the resulting 
distributions of $\log W_{2796}$ offsets with purple, magenta, red, and
brown histograms. 
Each histogram is generated from 2500 simulated $W_{2796} (A_{\rm G})$
measurements, and has been
normalized such that the peak value is equal to the peak of the
histogram showing the observed $\Delta \log W_{2796}$ distribution.


We demonstrate that these simulated datasets are
approximately normally distributed by performing a non-linear least
squares fit of a Gaussian to each, and show the best-fit curves with
thick colored lines.  We print the mean ($\mu$) and standard deviation
($\sigma$) of each sample above the corresponding plot panel.





\begin{figure}
\begin{center}
\includegraphics[angle=0,width=\columnwidth,trim=10 80 40
120,clip=]{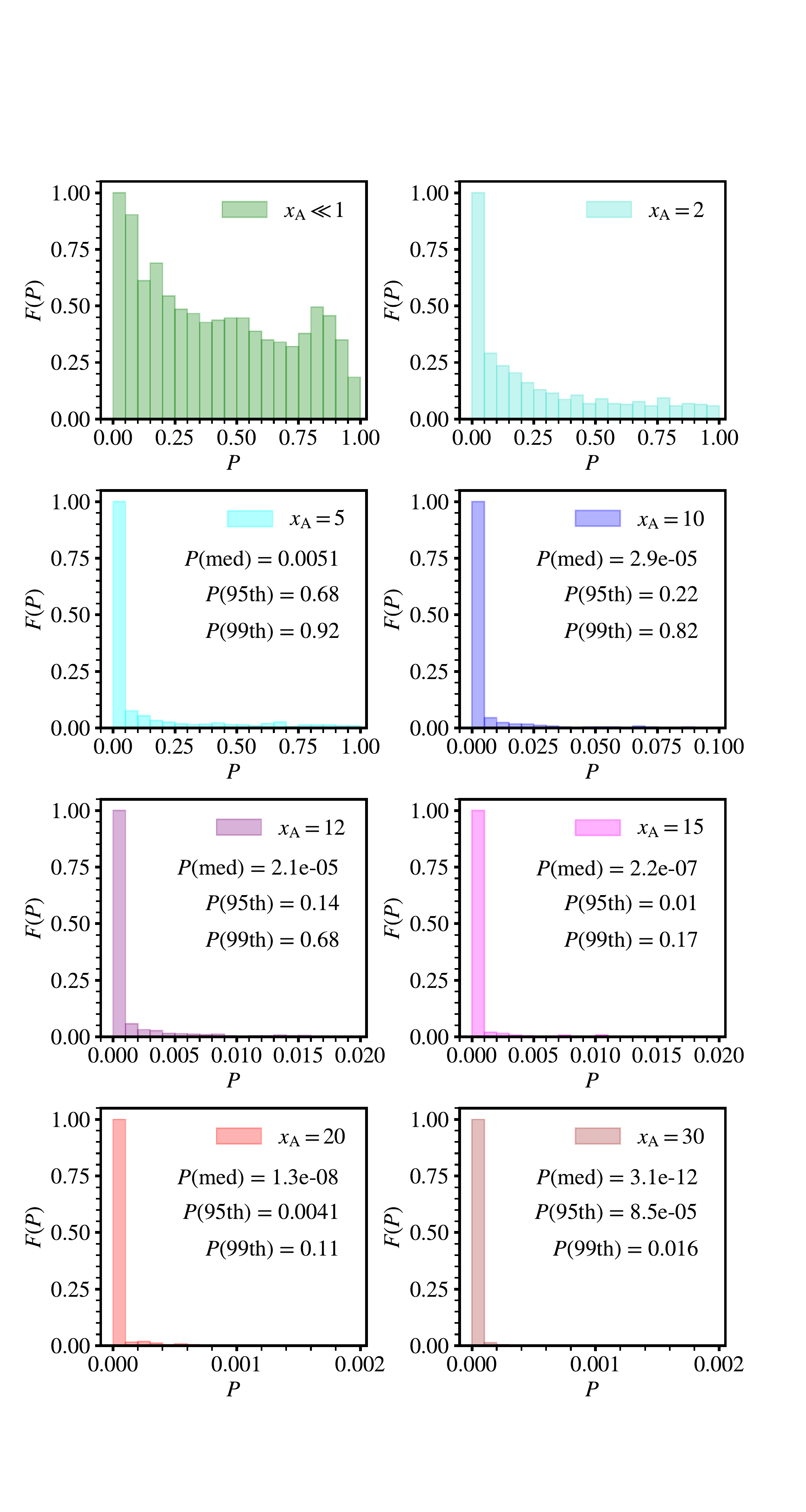}
\caption{Frequency distribution of $P$-values resulting from log-rank
  tests comparing the survival distributions of the observed $\Delta \log W_{2796}$ dataset and
  those of the
  simulated datasets with $x_{\rm A} \ll 1$ (green histogram), $x_{\rm
    A} = 2$ (turquoise histogram),
  $x_{\rm A} = 5$ (cyan histogram), 
  $x_{\rm A} = 10$ (blue histogram),
  $x_{\rm A} = 12$ (purple histogram), 
  $x_{\rm A} = 15$ (magenta histogram), 
  $x_{\rm A} = 20$ (red histogram), and $x_{\rm A} = 30$ (brown histogram).  The median and upper 95th- and 99th-percentile $P$-values
  are printed in the corresponding panels for $x_{\rm A} \ge 5$.
  $P$-values ruling 
out the null hypothesis are common for
  these latter distributions, pointing to statistically-significant
  inconsistencies between the observed and simulated $\Delta \log W_{2796}$ distributions.
\label{fig.primus_logranktest}}
\end{center}
\end{figure}

\begin{figure}
\begin{center}
\includegraphics[angle=0,width=\columnwidth,trim=15 20 15
15,clip=]{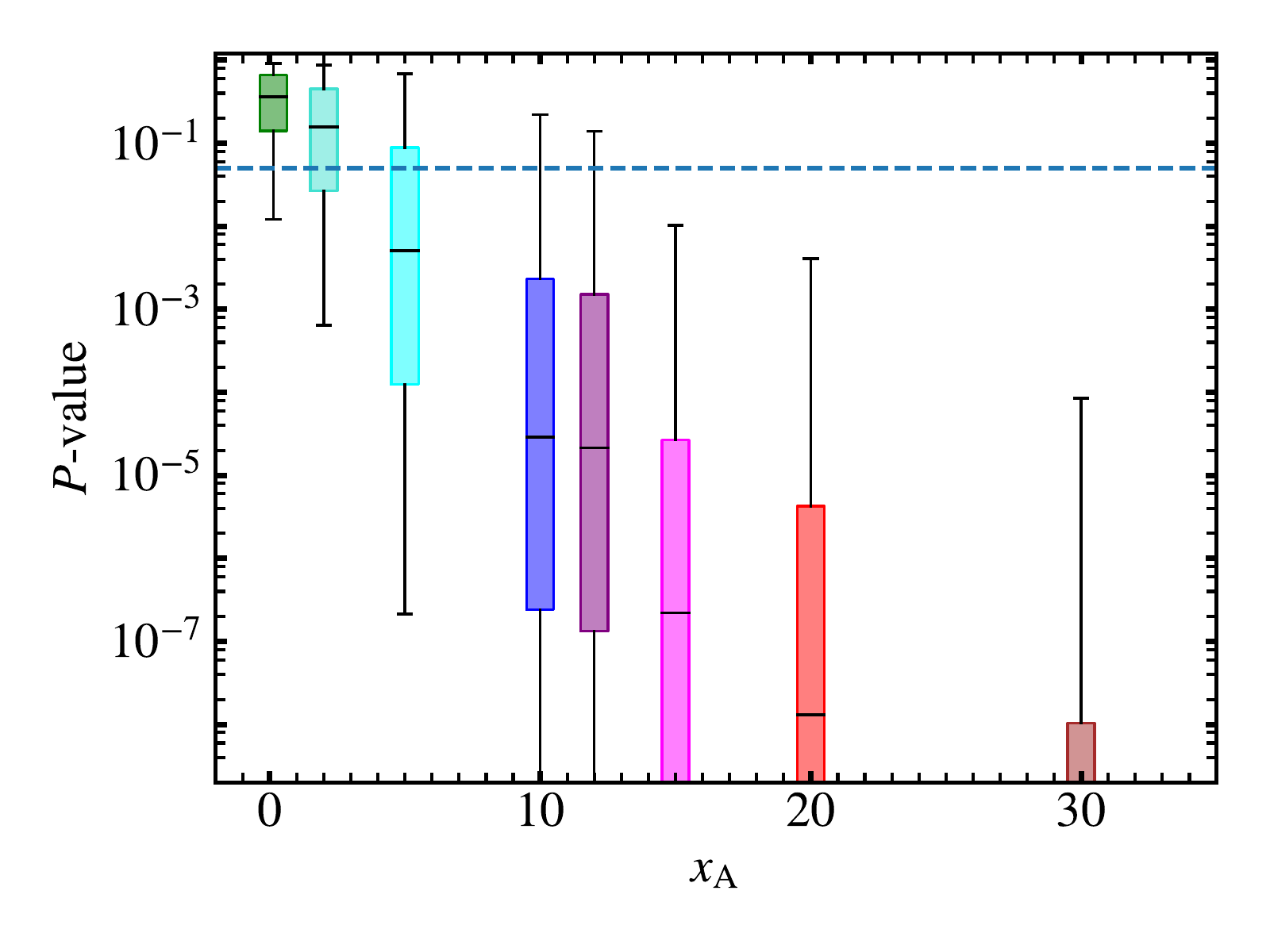}
\caption{Box-and-whisker plots representing each of the $P$-value
frequency distributions shown in Figure~\ref{fig.primus_logranktest}.
Each box extends to the upper and lower quartile values, with the
median indicated by the central black line.
The whiskers extend to the 5th- and 95th-percentile values.  The box colors are
the same as those used for the corresponding $x_{\rm A}$ values in
Figures~\ref{fig.primus_1dewdist} and \ref{fig.primus_logranktest}.
The dashed blue horizontal line indicates $P=0.05$.  Over 95\% of the
realizations adopting $x_{\rm A} = 15$ yield $P$-values below this limit,
allowing us to reject the null hypothesis that the observed and
simulated $\Delta \log W_{2796}$ distributions arise from the same
parent population.  Simulations with $10 \le x_{\rm A} \le 12$ also yield $P$
distributions dominated by values $\ll 0.05$, but in addition give rise to a
non-negligible fraction of realizations for which the null hypothesis cannot be rejected.
\label{fig.primus_summary_logranktest}}
\end{center}
\end{figure}


To test for consistency between the observed $\Delta\log W_{2796}$
distribution and each of these simulated datasets, we implement a
survival analysis using the Python package
\texttt{lifelines} \citep{Davidson-Pilon2018}.\footnote{Available at
  https://github.com/CamDavidsonPilon/lifelines.}
For each value of $x_{\rm A}$, we first perform a log-rank test comparing the survival distributions
of a Gaussian function having the corresponding values of $\mu$ and $\sigma$
 and the observed set of $\Delta \log W_{2796}$
values.  Here we include all measurements constraining the CGM of
star-forming f/g galaxies, including all upper  limits on $\log
W_{2796}$ (regardless of their value).\footnote{We
  also multiply both simulated and
observed datasets by $-1$, as the \texttt{lifelines} implementation of
the log-rank test handles right-censored data only.}  
The resulting $P$-values are printed above the appropriate panels in Figure~\ref{fig.primus_1dewdist}.

Then, to account for the magnitude of the uncertainty in each observed value
of $\Delta \log W_{2796}$, we consider both the uncertainty in
the best-fit fiducial relation  and the measurement
uncertainty in $W_{2796}$ ($\sigma_i$) at each value of
$R_{\perp, i}^{\rm corr}$.  First, we draw 1000 random sets of the parameters
$m_1$, $m_2$, and $b$ from the PPDF of our linear
\ion{Mg}{2}-absorbing CGM model
(as in \S\ref{sec.fiducial}).  We adopt the parameters in each of
these sets as fiducial parameters and recalculate $\log
\overbar{W}_{2796}$ accordingly.  For each set, we also generate a new
realization of the observed  $\log W_{2796}$ values, perturbing each detection by a
random draw from a Gaussian distribution having a standard deviation
equal to $\sigma_{i}$.  Finally, we perform the log-rank
test for each of these 1000 realizations, and record the resulting $P$-values.


We show the detailed 
distributions of these $P$-values in 
Figure~\ref{fig.primus_logranktest}.
 We summarize the
distributions in box-and-whisker form and compare them to our null
hypothesis rejection criterion ($P=0.05$) in 
Figure~\ref{fig.primus_summary_logranktest}.  
For datasets simulated with $x_{\rm A} < 5$, values of $P > 0.05$ are
quite common (see green and turquoise histograms), such that we fail to reject the null
hypothesis in these cases (i.e., we find no evidence that the observed
and simulated survival distributions differ).  For values of $x_{\rm A} \ge
5$ (for which $P$-values ${>}0.05$ are relatively infrequent), 
we print the median and upper 95th- and 99th-percentile $P$-values
in the corresponding panels of Figure~\ref{fig.primus_logranktest}.    

In detail, these results indicate inconsistency between the observed
and simulated $\Delta \log W_{2796}$ distributions with increasing
statistical significance as $x_{\rm A}$ increases from $10$ to
$30$.  Referring first to the direct comparison between the observed
set of $\Delta \log W_{2796}$ values (adopting the best-fit fiducial
CGM) and the simulated distributions, we reject the null hypothesis
wherever $x_{\rm A} \ge 10$. 
Considering the effects of the combined uncertainties in the fiducial
CGM model and in our $W_{2796}$ measurements, 
we find that the null hypothesis is ruled out in the majority of
realizations for 
$x_{\rm A} =5$ (with $P\rm (med) = 0.005$);
however, the log-rank test yields $P>0.05$ for $29\%$ of these
realizations.   This is reflected in
Figure~\ref{fig.primus_summary_logranktest} in the overlap of the
colored box at $x_{\rm A}=5$ showing the upper and lower quartile values of
the $P$-value distribution with the horizontal line showing $P=0.05$.
When $x_{\rm A} = 10$, the null hypothesis is ruled out for
$90\%$ of realizations, while for $x_{\rm A} = 15$, $P<0.05$ for all
but 22 of the 1000 Monte Carlo realizations of the observed dataset
(such that
  the whisker at $x_{\rm A}=15$ in
  Figure~\ref{fig.primus_summary_logranktest} ends below $P=0.05$).  
Thus, at the most conservative level, we
conclude that within the framework of this CGM model,
we must reject the null hypothesis for ratios $x_{\rm A} \ge 15$ with $95\%$
confidence.  

Moreover, given the frequency with which the
simulated $\Delta \log W_{2796}$ distribution for $x_{\rm A} = 10$
is demonstrated to be inconsistent with that observed, 
we also consider models having $10 \le x_{\rm A} < 15$
to be at least marginally inconsistent with our PRIMUS pairs dataset.

\subsection{A Limit on the Coherence Scale of $W_{2796}$ in the Inner CGM}\label{sec.result_coherence}

We may now use these constraints on the ratio of the projected areas
of our b/g galaxy beams to that of \ion{Mg}{2} absorbers ($x_{\rm A}$) to place a
limit on the projected area over which $W_{2796}$ does not
fluctuate.  We will refer to the square root of this area as the
``coherence scale'' or ``coherence length'' of \ion{Mg}{2} absorbers,
$\ell_{\rm A}$.

In \citetalias{GPG1}, we performed a detailed analysis assessing the spatial
extent of the rest-frame UV continuum emission arising from our b/g
galaxy sample (see Section 6).  We found that the vast majority of the
galaxies for which {\it HST} imaging is available in the rest-frame
optical (and which do not host bright AGN) have effective radii
$R_{\rm eff}(z_{\rm f/g}) > 2.0$ kpc at the redshift of the corresponding f/g galaxy
($z_{\rm f/g}$).  The median effective radius of this b/g galaxy
sample is $R_{\rm eff} (z_{\rm f/g})= 4.1$ kpc.  We also demonstrated that in
general, the half-light radii of bright galaxies at $z\sim0.5-1$ are very
similar in both the rest-frame optical and rest-frame UV, justifying
the assumption that the projected extent of our b/g galaxy sample as
measured in the rest-frame optical reflects the sizes of the continua
probing f/g \ion{Mg}{2} absorption.  

Adopting the size limit $R_{\rm eff} (z_{\rm f/g}) > 2.0$ kpc, such
that $A_{\rm G} > \pi (2.0^2)~\rm kpc^2$, our most conservative limit on
$x_{\rm A}$ ($x_{\rm A} = A_{\rm G}/A_{\rm A}$ is not $\ge 15$) requires
$A_{\rm A} > \frac{4.0\pi}{15}~\rm kpc^2$, such that $\ell_{\rm A} > 0.9$
kpc.  A somewhat less conservative constraint may be calculated by adopting
the median b/g galaxy size, yielding $A_{\rm G} > \pi (4.1^2)~\rm
kpc^2$, $A_{\rm A} > \frac{16.8\pi}{15}~\rm kpc^2$, and $\ell_{\rm A} > 1.9$ kpc.
Given that the b/g galaxies in our sample that have been imaged by {\it
  HST} range in size from $R_{\rm eff} (z_{\rm f/g}) = 1.0$ kpc to as large as
$R_{\rm eff} (z_{\rm f/g}) = 7.9$ kpc, we prefer this latter limit, and will adopt
it in the discussion that follows.  If we instead were to adopt the less
secure limit
$x_{\rm A} < 10$, the corresponding constraint on the
coherence length (again using $R_{\rm eff} (z_{\rm f/g}) = 4.1$ kpc) is $\ell_{\rm A} >
2.3$ kpc.  We expect that an expansion of  this galaxy pair
dataset and a larger sample of QSO-galaxy pairs suitable for
establishing the form of the ``fiducial'' CGM will eventually improve the
robustness of this larger limit.


Regardless of potential future developments, 
the foregoing analysis has placed a unique limit on the coherence
scale of the \ion{Mg}{2}-absorbing CGM within $\mrperp < 50$ kpc of
${\sim}L^*$ galaxies in the low-redshift universe ($\ell_{\rm A} > 1.9$ kpc).  
 This limit is fully consistent with those derived in previous
  work analyzing single low-redshift galaxy pairs with impact
  parameters $\mrperp < 50$ kpc (which have yielded $\ell_{\rm A} > 0.4$
  kpc; \citealt{Rubin2010,Diamond-Stanic2016,Peroux2018}). 
However, as noted above, this does not necessarily
imply that the material giving rise to absorption over this scale
must occupy a single structure subtending this area.  On the contrary,
there is a high degree of degeneracy in the physical locations and
velocities of ``clouds'' producing absorption of a given $W_{2796}$
even along a single pencil beam sightline.  We will consider the
implications of this constraint on the physical structure of
\ion{Mg}{2} absorption in the CGM in
detail in \S\ref{sec.interp}.

\section{Discussion}\label{sec.discuss}

\subsection{The Physical Nature of the \ion{Mg}{2}-Absorbing CGM}\label{sec.interp}

The power of QSO absorption line spectroscopy goes far beyond the
simple assessment of the incidence and distribution of the \ion{Mg}{2}
absorption strength ($W_{2796}$) in the halos of galaxies.  
High spectral resolution observations of QSO sightlines have
revealed the velocity and column density structure of these
absorbers in exquisite detail. Here we describe these constraints and
consider their implications for the interpretation of our limit on the
coherence length of $W_{2796}$ in the inner CGM.

\subsubsection{Relating Cool Structures and \ion{Mg}{2} Absorbers}

 The analysis of 
Keck/HIRES QSO spectroscopy probing ${\sim}23$ \ion{Mg}{2} absorbers at
$z\sim0.4-1.2$ by
\citet{ChurchillVogt2001} and \citet{Churchill2003} 
remains one of the most careful and germane in this context.
These authors performed Voigt-profile fitting of systems ranging in
strength from $W_{2796} \sim 0.3$ to $1.5$ \AA, finding that without
exception, the absorbers consisted of multiple Voigt profile
``components''.  In general, these components are kinematically narrow
with Doppler parameters $b_{\rm D} \sim5\mkms$ $-$ a velocity width 
that naturally arises from thermal broadening in gas at temperatures
$T\sim(3-4)\times10^{4}$ K \citep{Churchill2003}.  These individual, narrow
components  have frequently been attributed to structures  called
absorbing ``clouds'' in the literature
\citep{Churchill1999,ChurchillCharlton1999}, and we will continue to
use this term here.

In detail, \citet{Churchill2003} found that the number of these clouds
composing each \ion{Mg}{2} absorber  ($N_{\rm cl}$) ranged from two to
18, with  $N_{\rm cl}$
increasing approximately linearly with $W_{2796}$
(albeit with significant scatter).  A linear fit to their dataset
yielded a slope $= 0.058\pm0.004~\rm \AA~cloud^{-1}$ and intercept
$W_{2796} = 0.28$ \AA\ for $N_{\rm cl} = 1$.  
A few systems lying well above this trend were found to
be associated with 
heavily saturated \ion{Mg}{2} components arising from Damped
Ly$\alpha$ Systems (DLAs) or Lyman Limit Systems (LLSs),
and the authors noted that the number of components in these cases
could have been underestimated due to kinematic overlap or blending.  Finally, the
column density distribution function of all clouds in the absorber
sample was measured to a limiting column density of $\log
N($\ion{Mg}{2}$)\approx11.6~\rm cm^{-2}$, and was found to be
consistent with a power law $f(N) \propto N^{-\delta}$ with $\delta =
1.59 \pm 0.05$.  This implies that, for example, a cloud with column density
$\log N($\ion{Mg}{2}$)\approx12.5~\rm cm^{-2}$ occurs with more than an
order of magnitude higher frequency than a cloud with
$\log N($\ion{Mg}{2}$)\approx13.5~\rm cm^{-2}$.

These results are suggestive of a scenario in which \ion{Mg}{2}
absorbers, regardless of their total strength, are composed of
multiple, kinematically cold clouds superimposed along the same line
of sight.  The resultant \ion{Mg}{2} ``system'' will be stronger if
the sightline pierces a larger number of these clouds, and/or if there
is a large kinematic dispersion from cloud to cloud.  The
\citet{Churchill2003} sample does not include systems as strong as
many of the absorbers detected in our PRIMUS b/g galaxy sightlines (with $W_{2796} >
2$ \AA), and high-resolution spectroscopy of similarly strong absorbers tends
to yield broad, line-black profiles \citep{Mshar2007}.  While it is
difficult to demonstrate that such profiles arise from multiple narrow
components from analysis of \ion{Mg}{2} alone, examination of weaker
\ion{Fe}{2} transitions associated with the same systems generally
reveals a complex, multiple-cloud structure \citep{Mshar2007}.

\subsubsection{Exploration of Degeneracies in Component Structure and
  Velocity Dispersion}\label{subsubsec.degeneracies}

The assumption that \ion{Mg}{2} systems are indeed dominated by
kinematically cold clouds has important implications for
our interpretation of the coherence scale ($\ell_{\rm A}$) discussed in
\S\ref{sec.result_coherence}.  We emphasize that  $\ell_{\rm A}$
refers to the scale over which $W_{2796}$ does not fluctuate, rather
than a length scale associated with the cold clouds themselves.  These
quantities must be considered independently due to the degeneracy in the
relationship between $W_{2796}$, the number of clouds composing an
absorber ($N_{\rm cl}$), and the kinematic spread of these clouds. 
For instance, numerous saturated clouds all having a very similar radial velocity
(but that are physically quite distant from each other) can in
principle yield a much smaller $W_{2796}$ than a few clouds  with
velocities separated
by ${>}50\mkms$.

To better quantify the extent of this degeneracy, we simulate a suite
of \ion{Mg}{2} absorbers using code included in the Python package
\texttt{linetools}\footnote{Available at https://github.com/linetools.}.  To
simplify our experiment, we first assume that 
each cloud composing these absorbers has the same Doppler parameter
and column density.  
 We choose a Doppler parameter $b_{\rm D} = 5.4\mkms$
corresponding to the median value for the \citet{Churchill2003} sample,
and a column density $\log N($\ion{Mg}{2}$) =
13.0~\rm [cm^{-2}]$ toward the center of the column density
distribution of this sample (as shown in, e.g., Figure 4a of \citealt{Churchill2003}).  We
then generate line profiles for absorbers having between $N_{\rm cl} =
1$ and $20$ cold clouds.  For each system, the
velocities of the clouds are drawn at random from a Gaussian with a
full width at half maximum
$\Delta v_{\rm FWHM}= 150, 350$, or $550\mkms$.  These values sample the expected range in
the velocity dispersions of dark matter halos hosting galaxies with
$9.1 < \log M_*/M_{\odot} < 11.1$ at $z=0.5$ (assuming halo masses in
the range $11.3 \lesssim \log M_h/M_{\odot} \lesssim
13.3$; \citealt{MB2004,Moster2013}).  We generate 100 realizations of
each system (i.e., at a given $N_{\rm cl}$ and $\Delta v_{\rm FWHM}$),
and measure the resulting $W_{2796}$.  

\begin{figure}
\begin{center}
\includegraphics[angle=0,width=\columnwidth,trim=20 20 20
20,clip=]{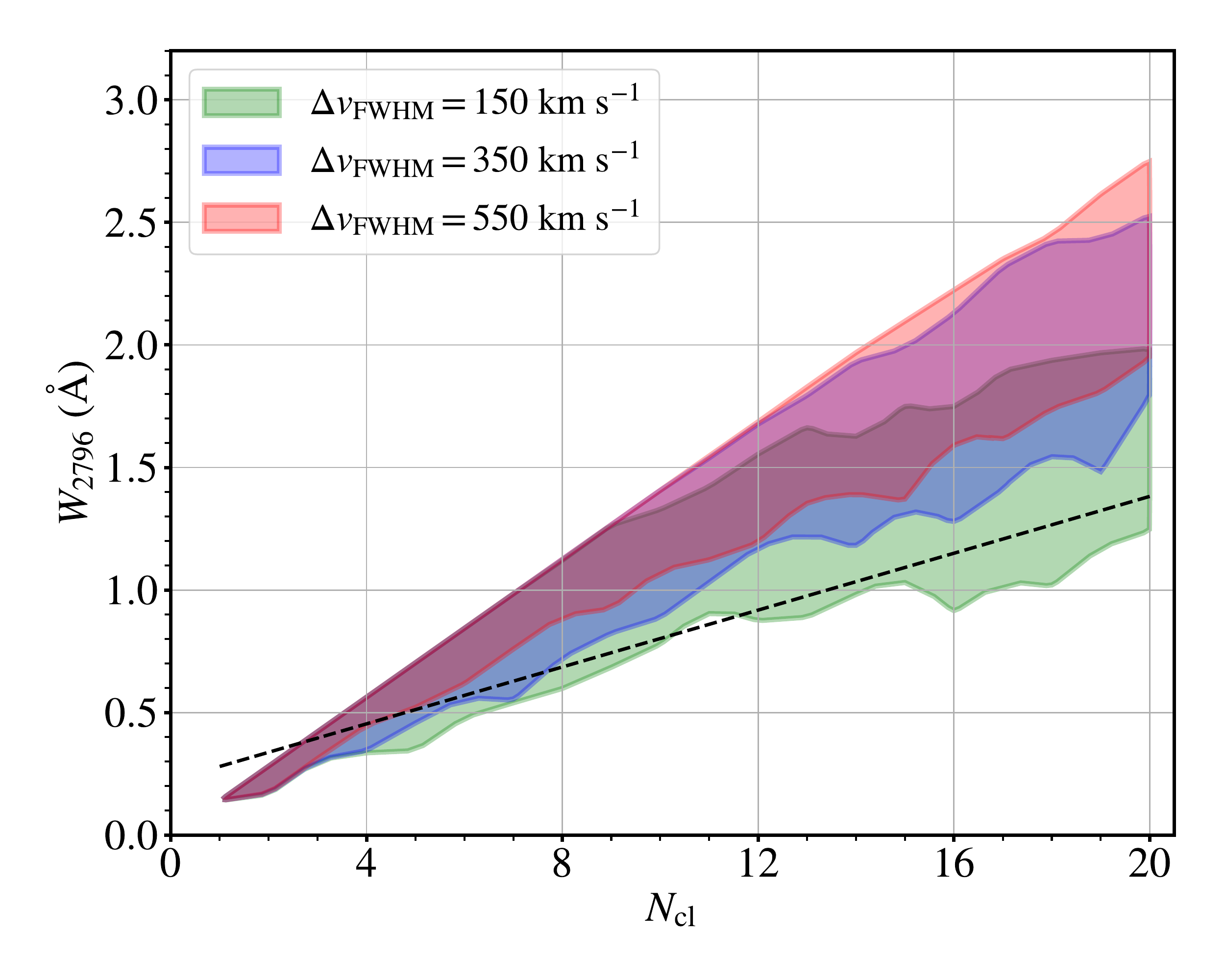}
\caption{Distribution of $W_{2796}$ measured from simulated \ion{Mg}{2} line
  profiles.  Each profile is composed of $N_{\rm cl}$
  kinematically-cold \ion{Mg}{2}-absorbing ``clouds''.  The clouds in
  every profile have velocities drawn at random from a Gaussian
  distribution with
  FWHM $ = \Delta v_{\rm FWHM}$.  For each value of $N_{\rm cl}$ and $\Delta v_{\rm
  FWHM}$ we generate 100 line profile realizations, and indicate the
resulting range in $W_{2796}$ with the green ($\Delta v_{\rm FWHM} =
150\mkms$), blue ($\Delta v_{\rm FWHM} = 350\mkms$), and red ($\Delta
v_{\rm FWHM} = 550\mkms$) contours.  The dashed line shows a linear
fit to the number of clouds vs.\ $W_{2796}$ for the sample of strong
\ion{Mg}{2} absorbers discussed in 
\citet{Churchill2003}.
\label{fig.primus_vdisp_ncl}}
\end{center}
\end{figure}

The ranges in these $W_{2796}$ values are indicated by the colored
contours in Figure~\ref{fig.primus_vdisp_ncl}.  The outlines of the
contours are placed at the minimum and maximum $W_{2796}$ for every
value of
$N_{\rm cl}$.  For comparison, the linear fit to the $N_{\rm
  cl}-W_{2796}$ relation for the \ion{Mg}{2}
absorber sample discussed in \citet{Churchill2003} is indicated with a
black dashed line.  The intercept of this relation at $N_{\rm cl} = 1$
is slightly higher than the $W_{2796}$ of our $N_{\rm cl} = 1$ line
profiles, suggesting that the weakest absorbers in the
\citet{Churchill2003} sample have a Doppler parameter or column
density slightly greater than the values we have chosen for each individual ``cloud''.  However, this fit otherwise
corresponds quite closely to the lower envelop of $W_{2796}$ values
exhibited by profiles with
$\Delta v_{\rm FWHM} = 150\mkms$.  The overall higher $W_{2796}$
distribution of our simulated sample is likely driven by the relatively
low velocity spread of clouds giving rise to the absorbers in the
\citet{Churchill2003} sample: most of these systems are produced by
clouds with relative velocities $\Delta v <100\mkms$, and the system
with the highest velocity spread in the sample has only $\Delta v \sim
400\mkms$.  Figure~\ref{fig.primus_vdisp_ncl} demonstrates that
systems with velocity dispersions $\lesssim 150\mkms$ are unlikely to yield
$W_{2796} > 2$ \AA; and indeed, the maximum $W_{2796}$ among the
\citet{Churchill2003} absorbers is ${\sim}1.5$ \AA.


Figure~\ref{fig.primus_vdisp_ncl} further demonstrates that there are
numerous physical scenarios that may give rise to absorbers with a
given $W_{2796}$.  For instance, an absorber having $W_{2796}$ close to
the median
value for the PRIMUS pair sightlines shown in Figure~\ref{fig.primus_galfid}, $W_{2796} \approx 0.7$
\AA, can arise from a system of $N_{\rm cl} \sim 5-6$ clouds with a
large velocity spread, or from $N_{\rm cl} \sim9-10$ clouds with a low
velocity dispersion such that the individual cloud line profiles
overlap in velocity space.
The range of $N_{\rm cl}$ values that
can plausibly yield a given absorber strength broadens as $W_{2796}$
increases.  
EWs as large as the maximum of the observed distribution  ($W_{2796} =
2.6$ \AA) are only exhibited by
 the simulated systems having
$N_{\rm cl}\sim19-20$ and a ``maximal'' velocity spread. 
However, our choice to limit $N_{\rm cl}$ to $\leq20$
is not driven by any physical constraint, and such strong systems
could easily result from absorbers with $N_{\rm cl} > 20$ and with
less extreme kinematics.  


\begin{figure*}
\begin{center}
\includegraphics[angle=0,width=7in,trim=10 110 15
30,clip=]{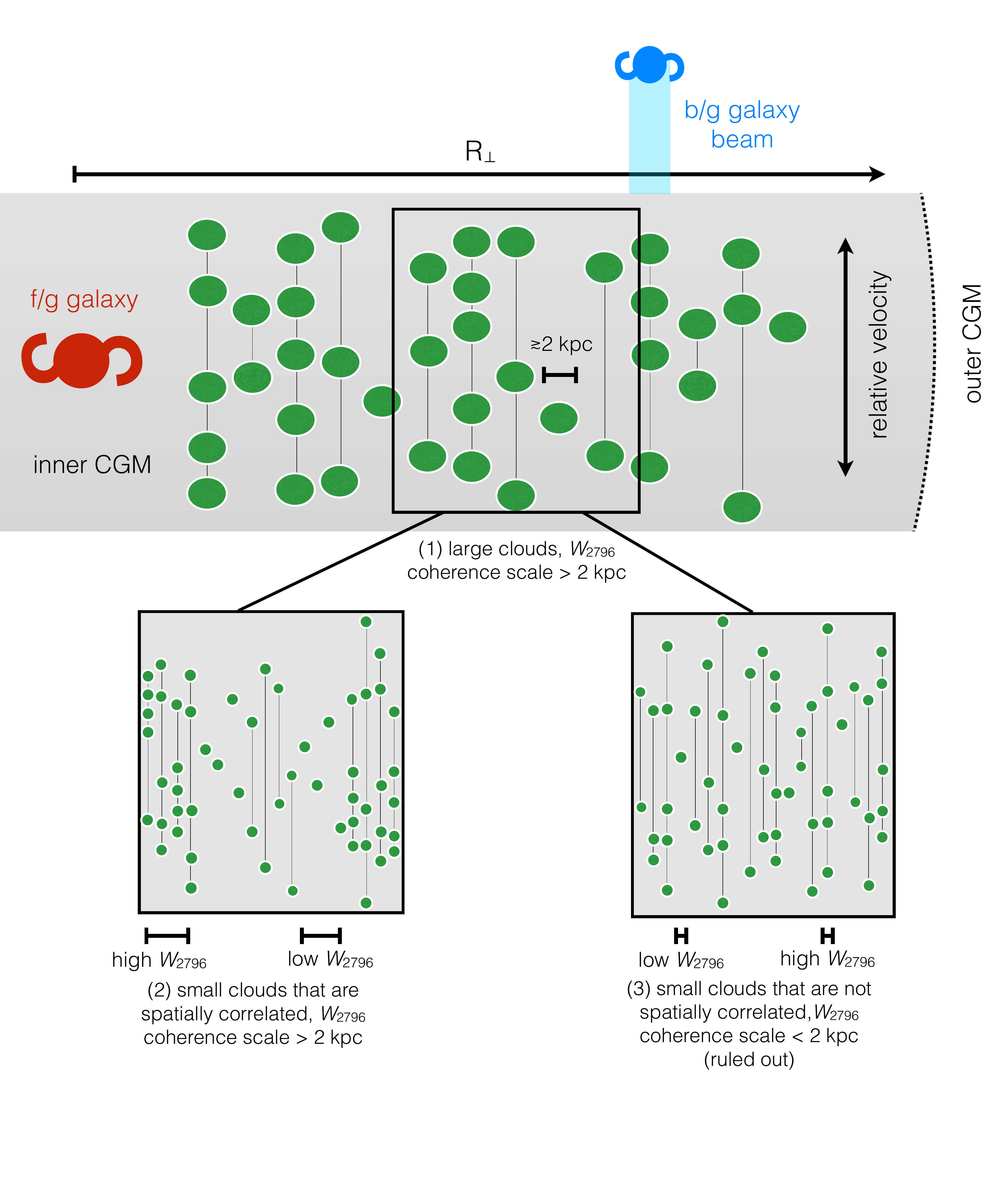}
\caption{Several scenarios for the physical
  distribution of \ion{Mg}{2}-absorbing structures in the inner CGM
  ($\mrperp < 50$ kpc).    Panel {\bf (1)} shows a slice through this
  inner halo region (gray) centered around a host galaxy indicated in red.
Each green circle or oval represents a
  distinct structure or cloud, which we assume gives rise to saturated (or
  nearly saturated)
  \ion{Mg}{2} absorption (having $\log N$(\ion{Mg}{2}) $> 13.0~\rm [cm^{-2}]$) with a
  narrow velocity dispersion ($b_{\rm D}\sim 5-10\mkms$).  These
  structures are placed along the vertical axis according to their mean velocity relative to the
  host galaxy.
Sightlines that
  pass through several clouds (as indicated by the thin vertical black lines) give rise to large
  $W_{2796}$, while sightlines passing through only a single cloud
  yield $W_{2796}\sim 0.2$ \AA.  The rough size of one of our b/g galaxy
  beams is indicated by the blue symbol at the upper right. 
 In panel
  {\bf (1)}, each individual cloud fully covers the beam of the b/g
  galaxy, such that the
  velocity structure within the beam is coherent.  
In panel {\bf (2)}, the physical scale
over which $W_{2796}$ changes is
  similar to that of the b/g galaxy beam; however, the
  high-$W_{2796}$ regions contain numerous, smaller clouds with
  incoherent velocities.  
The $W_{2796}$ distribution observed toward the
PRIMUS b/g galaxy sightlines
  is consistent with both of these scenarios.  
Our analysis rules out the scenario shown in panel {\bf (3)}, in
which the $W_{2796}$ varies on scales smaller than that subtended by
our b/g galaxy beams.  
\label{fig.primus_discuss}}
\end{center}
\end{figure*}

\subsubsection{Interpretation of the $W_{2796}$ Coherence Scale}\label{subsubsec.interp_coherence}

In light of these considerations, 
we now return to the 
interpretation of our constraint on the coherence scale of $W_{2796}$.
Our analysis requires that $\ell_{\rm A} > \cllength$ kpc, or $A_{\rm A} >
\clarea~\rm kpc^2$.
Thus, if we consider 
a contiguous ``patch'' of the CGM with a projected area $A_{\rm A}$,
the corresponding three-dimensional column of the CGM must be
populated with absorbing structures that give rise to the same
$W_{2796}$ across this entire area.  These structures, however, may
have a wide variety of configurations.  

We present cartoons
illustrating three such configurations in
Figure~\ref{fig.primus_discuss}.
Here, the topmost cartoon diagrams the layout of the ``inner'' CGM.  The
location and approximate scale of the host galaxy is indicated with
the red symbol.  The gray region represents a slice through the CGM
with impact parameter increasing to the right.  
We adopt the assumption  that the
absorption is in any case produced by numerous, kinematically-narrow clouds as
suggested by \citet{Churchill2003}, and represent these structures in
green.
We place these clouds
along the vertical axis according to their velocity relative to the
f/g host galaxy.  The projected extent of the beam toward a typical 
b/g galaxy is indicated in blue. 


The simplest scenario that can
yield the required coherence invokes clouds which themselves have
projected areas $> \clarea~\rm kpc^2$ as illustrated in
panel (1) of Figure~\ref{fig.primus_discuss}.
Numerous clouds observed along a given beam yield the strongest
absorbers, while beams absorbed by only a few clouds yield weak
absorption.  Based on the computations discussed in
\S\ref{subsubsec.degeneracies}, 
the weakest
absorption observed toward our b/g galaxy sample is sufficiently
strong that it likely arises from ${\sim}3-5$ such clouds, 
while the strongest absorbers would be composed of ${\gtrsim}20$
of these extended clouds.  (We have reduced these numbers for the
purposes of clarity in the figure.)



Alternatively, each of these kinematically-cold clouds could instead subtend a
much smaller area; e.g., several square parsecs.  The weakest absorbers in
our b/g galaxy sample would in this case arise from many thousands of
clouds (as would be required to cover an area $>1~\rm kpc^2$).
However, they would be distributed such that the sightline through each
area element of the patch
($dA_{\rm A}$) intercepts only ${\sim}3-5$ clouds.  
In regions of the CGM with the strongest observed absorption, many thousands
of clouds would again be required, and here each element $dA_{\rm A}$
would be shadowed by ${\gtrsim}20$ clouds.  Note that the clouds along
the sightline through any given subpatch $dA_{\rm A}$ may be completely
physically distinct from those clouds within a neighboring
subpatch.  They may have different physical locations and velocities
within the halo, and the number of clouds from subpatch to subpatch
may also vary as long as the cloud velocity dispersion yields a
consistent $W_{2796}$ throughout the patch.  

This second scenario  is illustrated in panel (2) of
 Figure~\ref{fig.primus_discuss}.  
Here, each cloud subtends a smaller projected area than in panel (1)
(they have diameters approximately $1/3$ the width of the b/g galaxy
beam).  The clouds exhibit significant variations in their relative
velocities from sightline to sightline.
However, the numbers of clouds along different   
sightlines are spatially correlated, with a ``correlation length'' at
least a large as $\ell_{\rm A} > \cllength$ kpc.  In other words, regions of
the halo having numerous, small clouds along a given sightline 
tend to have projected areas $> \clarea ~\rm kpc^2$.  Low-$W_{2796}$
regions arising from the absorption of only a few of these small
clouds also tend to extend over areas larger than the clouds
themselves.  As noted above, the pencil-beam sightlines  within
a given region of coherence need not all pierce the same number of
clouds (as  is shown in the figure).  We have illustrated the scenario this
way for simplicity's sake; however, our observations require only that
the \emph{combination} of $\Delta v$ and $N_{\rm cl}$ yield similar
$W_{2796}$ within a region $A_{\rm A}$.

Panel (3) of Figure~\ref{fig.primus_discuss} illustrates a final
scenario: one which is ruled out by our observations.  Here, the individual
\ion{Mg}{2}-absorbing structures have small sizes as in panel (2).
However, the number of clouds varies significantly from one sightline
to the next, such that $W_{2796}$ varies on projected spatial scales
much smaller than the b/g galaxy beam.  Similar variations could also arise
from significant changes in $\Delta v$ from sightline to sightline.  
Either mechanism is unlikely given the results of the analysis
described above.

In summary, on a phenomenological level, our measurements point to
either (1) \ion{Mg}{2}-absorbing
structures that extend over projected areas $>\clarea~\rm kpc^2$, or
to (2)  a
spatial correlation in the numbers of these clouds and their velocity
dispersion with a correlation length $> \cllength$ kpc.  
These two types of configurations may plausibly arise from a variety
of physical scenarios; e.g., extended, cool streams of inflowing gas,
or numerous cold clouds embedded in a rapidly-expanding and
kinematically-complex outflow.  
We also note here that echelle-resolution spectroscopy of
  extended b/g sources as presented in \citet{Diamond-Stanic2016} has
  the potential to differentiate between these scenarios, as it
  is sensitive to both velocity structure and variations in covering
  fraction across the beam.  Indeed, these authors report a covering
  fraction of unity  for both strong and weak \ion{Mg}{2} components at
 $\mrperp = 27$ kpc over a beam $\approx 0.4$ kpc in radius, lending
 support to scenario (1) above.
We will further consider the physical
implications of our  findings and these scenarios  in \S\ref{subsec.thermo}.




\subsection{Complementary Constraints on the Coherence of \ion{Mg}{2} Absorption}\label{subsec.complementary_constraints}

Our dataset is not the first to offer constraints on the sizes
of cool, photoionized structures in the CGM.  Spectroscopy of
gravitationally-lensed QSO sightlines yielded the first such
measurements beginning in the 1980s
\citep[e.g.,][]{Young1981,WeymannFoltz1983,Foltz1984}.  
Photoionization modeling provides yet another assessment of absorber
size, albeit indirect.  Finally, spectroscopy of UV-bright stars
or AGN probing cool gas in the halo of our own Galaxy constrains
 its absorption coherence over a broad range of length scales
\citep[e.g.,][]{Smoker2015}. 
 Here we briefly review the findings of these
experiments, and consider how they complement the present experimental
approach.

\subsubsection{The Inner CGM of the Milky Way}

A key testing ground for the coherence of circumgalactic material
is the Milky Way itself, in which the extent of cool gas clouds can be
viewed directly in 21 cm emission.  The largest recent  neutral
hydrogen surveys have been conducted with single dish radio telescopes
with spatial resolutions $4-36\arcmin$
\citep[e.g.,][]{Putman2002,Kalberla2005,Peek2011}, corresponding to
sizes of ${\sim}25-210$ pc at a distance of 20 kpc.  This imaging has
revealed a great variety of structures, with those at the highest
velocities (i.e., High Velocity Clouds, or HVCs) tending to be located beyond ${\sim}5$
kpc from the disk \citep{Wakker2001}.  The most massive of these
structures are organized into extended cloud complexes or streams
(e.g., the Magellanic Stream; Complex C) with physical dimensions of
several to several tens of kpc
\citep[e.g.,][]{Thom2008,DOnghiaFox2016}.  There is, in addition, a
substantial population of ``compact'', isolated HVCs
 with somewhat smaller sizes
(${\sim}5-15\arcmin$, with a median physical size of ${\sim}10$ pc
assuming a distance of 10 kpc; \citealt{Putman2002,Saul2012}). 
Interferometric imaging of a handful of neutral hydrogen cloud complexes in
the halo has revealed that they are
composed of numerous dense clumps only ${\sim}30-50$ pc across
\citep{Richter2005,BenBekhti2009}, suggesting that such fine
substructure may be a common feature of the HVC population as a whole.

However, the ionized gas associated with these clouds is known to
be significantly more extended than the neutral component
\citep{LehnerHowk2011}.  Furthermore,  the \ion{H}{1} velocities of
HVCs typically vary by $< 5\mkms$ over much of their projected
surface \citep{Wakker2008}, suggesting that if the associated
low-ionization absorption were probed along multiple close sightlines it
would exhibit minimal velocity shear.  In principle, analysis of
 the coherence of low-ion absorption toward  halo stars
or AGN 
       \citep[e.g.,][]{Andre2004,vanLoon2009,Nasoudi-Shoar2010,vanLoon2013,Smoker2015} 
has the greatest
potential of any of the techniques discussed here to reveal the
detailed density structure of circumgalactic clouds, and such
efforts are ongoing (J.~K.~Werk et al., in preparation).


\subsubsection{Gravitationally-Lensed QSOs}\label{subsubsec.lensing}

Gravitationally-lensed images of a single QSO can  provide
powerful constraints on the coherence of the intervening gas.
Depending on the properties of the lensing system and its
configuration relative to the source, the QSO images may appear
separated over a wide range of physical scales extending to several
tens of kiloparsecs.  The assembly of a large sample of such systems can
therefore offer much stronger leverage on the sizes of f/g absorbers
than 
the beams of unlensed b/g galaxies.  Lensed
QSOs are also often much brighter than the brightest b/g galaxies 
that may be selected for studies such as this, and hence permit high-S/N, high spectral resolution observations
probing the detailed kinematic coherence of the material.

Early spectroscopic studies of gravitationally-lensed QSOs focused
primarily on the size scales of \lya and \ion{C}{4} absorbers at $z\sim1-2$ 
\citep[e.g.,][]{Young1981,WeymannFoltz1983,Foltz1984,Smette1992}.
Using spectroscopy of QSO images
with projected separations ranging from $\mrperp \sim1~h_{50}^{-1}$ kpc to
$35 ~h_{50}^{-1}$ kpc, these works emphasize the strong similarity between
the equivalent widths of these species observed in adjacent
sightlines.   
In one of the first works to focus on constraining the coherence of low-ion metal
absorption 
using this technique, 
\citet{Smette1995} reported the identification of five \ion{Mg}{2} absorbers along 
doubly-lensed QSO
sightlines separated by
${\sim}10-25~h_{50}^{-1}$ kpc.  None of the systems having $W_{2796}< 0.4$
\AA\ were detected in more than one sightline. 
The two stronger systems (with $W_{2796} = 0.5-1.0$ \AA) yielded
absorption which varied by $\sim30-40\%$ from one sightline to the
other, suggestive of a scenario in which both sightlines probe the
``inner'' CGM of an associated intervening halo (at $\mrperp < 50$ kpc). 

More sophisticated analyses, made possible with the advent of {\it HST} and
high-resolution spectrographs on 10m-class telescopes, have 
gradually bolstered evidence for a picture in which weak \ion{Mg}{2}
absorbers are composed of clouds extending over less than a
kiloparsec, whereas stronger absorbers ($W_{2796} \gtrsim 1$ \AA) 
are coherent over larger scales \citep{Monier1998,Rauch1999,Petitjean2000,Rauch2002,Churchill2003lens,Ellison2004}.
For example, in their study of {\it HST}/FOS spectroscopy of the four
sightlines to the Cloverleaf QSO at $z\sim2.54$ (probing 
scales of ${\sim} 1-6~h^{-1}$ kpc), \citet{Monier1998} presented a detailed comparison
of the EWs across every sightline for the three intervening
systems with detected low-ionization metal absorption.  
They found that most of the
species with EW $\gtrsim0.5$ \AA\ exhibited absorption along all four sightlines,
and furthermore that in the strongest of these systems, the low-ion
EWs are very similar from sightline to sightline.  

Following this study, \citet{Rauch1999,Rauch2001,Rauch2002} 
presented a series of works
leveraging high-S/N spectroscopy obtained with Keck/HIRES
to examine the spatial structure of gas clouds on sub-kiloparsec
scales.  These authors  introduced
the term ``coherence length'' to describe the distance over which
there are significant changes in the physical parameters of a cloud
(e.g., column density or projected velocity; \citealt{Rauch2001}).  
We note that this usage differs from our intended meaning 
in the foregoing text: we have invoked this phrase to describe a lack of variation
in EW, rather than a significant variation in cloud physical parameters.\footnote{As
discussed in \S\ref{subsubsec.interp_coherence}, coherent $W_{2796}$ does not imply coherent
velocity structure.}
\citet{Rauch1999} discussed a single system at
$z\sim3.54$ along sightlines separated by only $26~h_{50}^{-1}$ pc.
Variations in the column densities of \ion{C}{2} and
\ion{Si}{2} between the two sightlines by factors of $\sim2-10$ and
velocity offsets of $\sim10\mkms$ were uncovered, suggesting that the
cool structures are composed of numerous tiny
``cloudlets''.  \citet{Rauch2002} then presented similar observations of
three additional systems giving rise to \ion{Mg}{2} and/or \ion{Fe}{2}
absorption at $z\sim0.5-1$ along three sightlines separated by ${\sim} 0.2-0.7~
h_{50}^{-1}$ kpc.  The strongest system ($W_{2796} = 1.2$ \AA\ at
$z=0.566$) had a qualitatively similar component structure across
all of these sightlines, although the precise velocity centroids were
observed to shift slightly, and most of
the components exhibited column density variations of up to 1.5 dex.  
The components comprising the weaker systems, on the other hand, were
frequently not detected along one or two of the sightlines.   


The more recent work of 
\citet{Chen2014} is one of very few studies to use lensed QSO sightlines
to probe within $\mrperp < 50$ kpc of bright f/g galaxies whose redshifts were
identified {\it a priori}.  These authors targeted a
quadruply lensed
QSO at $z \sim 1.7$ lying close behind two ${\gtrsim} L_*$
galaxies at $z=0.42$ and $z=0.78$.  Magellan/MagE and MIKE spectroscopy of the
four sightlines, each separated by $\sim5-10$ kpc 
at the redshifts of the foreground halos,
revealed ubiquitously strong ($W_{2796} \gtrsim 0.5$ \AA) absorption. 
In particular, the $z=0.42$ halo exhibited notably coherent velocity
profiles with a velocity shear of only $\Delta v \sim 20\mkms$ across
the sightlines.  The $z=0.78$ halo, on the other hand, gave rise to
qualitatively distinct \ion{Mg}{2}  profiles with line widths
differing by up to ${\sim}180\mkms$ and velocity centroids offset by
${\sim}90\mkms$. 
 The authors considered several physical origins for
the absorption, finally concluding that the properties of infalling
gas streams 
are most consistent with the observed line profiles (and more so than the
putative properties of an extended, rotating disk or biconical galactic winds).
It is suggested that the overall coherence of the profiles would arise
from streams $\gtrsim10$ kpc in width, and that turbulent motion
associated with the streams must contribute to the velocity centroid
offsets between sightlines.  
\citet{Zahedy2016} have also made use of lensed QSO sightlines to probe
the inner CGM, targeting the 
halos of three ${\gtrsim}L_*$ elliptical galaxies at  $\mrperp< 15$ kpc.
In the most spectacular of these systems, the two sightlines of a double lens, separated by ${\sim}8$ kpc, 
uncovered remarkably similar (and
complex) velocity profiles with an overall velocity offset of
${\sim}350\mkms$.  The authors liken this high level of coherence to that
observed in the \citet{Chen2014} study, while also noting the
difficulties in differentiating between prospective physical origins
for the gas.

In summary, spectroscopy of gravitationally-lensed QSO sightlines 
have begun to reveal the detailed spatial structure of low-ionization
absorption in a wide range of environments.  
Structures giving
rise to relatively weak absorption
($W_{2796} \lesssim 0.4$ \AA) tend to exhibit substantial variations in
EW and/or column density and velocity on scales less than ${\sim}1-2$ kpc 
\citep[e.g.,][]{Monier1998,Petitjean2000,Rauch2002,Ellison2004,RogersonHall2012}.  We note that such
systems are relatively rare in the inner CGM of ${\sim}L_*$ galaxies at $z
< 1$,
occurring along only four of the 22 QSO sightlines passing within $\mrperp
< 30$ kpc of the galaxy samples discussed in \citet{Chen2010} and \citet{Werk2013}
(see also Figure 11 of \citetalias{GPG1}).  Even at 30 kpc $< \mrperp <
$ 50 kpc, $W_{2796}<0.4$ \AA\ absorption occurs in only 14 of the 32
sightlines in these samples.  Stronger systems, in contrast, tend to
yield strong absorption and similar velocity structure in lensed QSO sightlines separated by
$\gtrsim1-5$ kpc \citep{Monier1998,Rauch2002,Chen2014}.
These results are  at least qualitatively consistent with our
finding that the EW of \ion{Mg}{2} absorbers observed along our b/g
galaxy sightlines does not vary on length scales $< \cllength$ kpc.  
Larger samples of bright, lensed QSOs will be needed to perform a
more quantitative test for the consistency of all of these constraints
(from, e.g., the Hyper Suprime-Cam Survey; \citealt{More2017},  or
VST-ATLAS; \citealt{Schechter2017}).  

Moreover, these studies may also
in principle be used to differentiate between scenarios (1) and (2) as
described in Figure~\ref{fig.primus_discuss} and
Section~\ref{subsubsec.interp_coherence}.
At present, the coherence of the CGM structures reported in
\citet{Chen2014} are evocative of the scenario diagrammed in panel
(1) of this figure, and tend to disfavor scenario (2).  We expect that
expanded absorption line studies using both lensed QSO and b/g galaxy
spectroscopy will be crucial to ruling out and/or refining these models.


\subsubsection{Photoionization Modeling}\label{subsubsec.photoionization}

The sizes of cool circumgalactic gaseous structures are also
manifest in the ionization state of the gas.  The thickness of
an absorbing cloud affects the extent to which it is penetrated by
ionizing radiation, which in turn affects the cloud's ionization
fraction \citep[e.g.,][]{Bergeron1986,DonahueShull1991,Ferland1998}.
Thus, under the assumption of a particular ionizing radiation field,
cloud configuration, and metal abundance pattern, measurements of the
column densities of rest-frame UV metal transitions spanning a range of
ionization states can yield an estimate of the cloud size
\citep[e.g.,][]{ChurchillCharlton1999, Rigby2002,LanFukugita2017}.


Over the last several years, this technique has frequently been
invoked in the context of $z\sim2-3$ CGM studies, for which many of the
relevant transitions are accessible in the optical.   
\citet{Crighton2013} analyzed absorption detected at $\mrperp = 58$ kpc
from a bright, star-forming galaxy at $z = 2.44$.  Their high-S/N, high spectral resolution coverage of the Lyman
series permitted very precise constraints on the neutral hydrogen
column density in this system.  The properties of one particularly
strong velocity component of this absorber (having $N($\ion{Mg}{2}$)=
10^{13.2}~\rm cm^{-2}$), together with their assumption of a
\citet{HM2012} ionizing background and solar abundance ratios, 
strongly suggested a cloud thickness $\lesssim3$ kpc.

Following this work, \citet{Crighton2015} reported on the detection of
a partial Lyman-limit system at $\mrperp = 50$ kpc from a ${\sim}0.2
L_*$ galaxy also at $z \sim 2.5$.  Here, the authors carefully treated
the systematic errors
associated with photoionization modeling,
introducing a variable slope to the input radiation field.  
They  also used the MCMC technique to calculate marginalized probability
distributions for each model parameter.  Their analysis constrained the cloud
thicknesses to be ${\sim}100-500$ pc, with typical uncertainties less than
a factor of $\sim2.5$.  
Similarly small (or even smaller) cloud sizes were recovered for the strong
absorption systems detected in the circumgalactic environments of
$z\sim2$ QSOs from analysis of the associated \ion{C}{2}* $\lambda
1335$ transition \citep{QPQ3,Lau2016}.

While such sub-kiloparsec or sub-parsec clouds are apparently typical
of the $z>2$  CGM, photoionization modeling of the CGM at lower
redshifts (now possible with the advent of {\it HST}/COS) suggests that cool clouds exhibit a much wider range of
sizes at the current epoch \citep[e.g.,][]{Stocke2013}.
\citet{Werk2014} performed photoionization modeling of the CGM
absorbers around a sample of $z\sim0.2$, ${\sim} L_*$ galaxies
observed as part of the COS-Halos survey
\citep{Tumlinson2011,Tumlinson2013,Werk2012,Werk2013}, 
leveraging a dataset containing
44 close ($\mrperp < 160$ kpc) sightlines.  Among this
sample, 33 sightlines exhibited low- or intermediate-ionization
absorption, permitting tight constraints on the ionization state and
metallicity of this subset.  The
resulting values of $N_{\rm H} / n_{\rm H}$ range between 0.1 and 2000
kpc with uncertainties of several orders of magnitude. 
 However, those absorbers with the strongest
\ion{Mg}{2} (i.e., $N($\ion{Mg}{2}$)> 10^{13.5}~\rm cm^{-2}$) tend to
also  have particularly well-constrained neutral hydrogen column densities from
fitting of the damping wings of the \lya profile, such that the
corresponding cloud sizes could be determined to within $\pm 1$ dex.
Moreover, these systems, which we suggest are close analogs to the
strong \ion{Mg}{2} absorbers we observe in the inner CGM in the
present sample, have overall larger sizes of ${\sim}7-100$ kpc
(J.~K.~Werk, private communication).

All together, we view this growing body of photoionization modeling
analyses as broadly consistent with the general picture supported by
the lensed QSO studies discussed in \S\ref{subsubsec.lensing}: weak,
low-ionization absorbers, and/or those at high redshift ($z > 1$; see
also \citealt{LanFukugita2017}), tend
to yield cloud thicknesses less than a kiloparsec (or have sizes which
are weakly constrained), whereas stronger
systems at $z<1$ residing close to a massive galaxy exhibit
thicknesses over ${\sim} 5-100$ kpc scales.  These results must of
course be considered with the caveat that they are subject to
substantial systematic uncertainties (e.g., in the strength and shape
of the  ionizing radiation field, the cloud metal abundance pattern,
or the
cloud geometry).  In particular, 
\citet{Stern2016} recently demonstrated that a cloud structure
allowing for multiple gas densities (rather than a constant-density
slab as assumed by \citealt{Werk2014}) can more closely match the full
range of absorption strengths measured in the COS-Halos dataset, and would
yield much smaller \ion{Mg}{2} absorber sizes ($\sim50$ pc;
\citealt{Stern2016}).  However, there is little {\it a priori}
observational evidence favoring this more complex geometry over the
simpler structures assumed in most previous studies.  Again, larger
samples of absorbers detected along lensed QSO sightlines will be
useful for differentiating among these models.



\subsection{The Significance of Our Modeling Assumptions}\label{subsec.assumptions}

Our analysis has relied on a number of
significant assumptions regarding the nature of the
\ion{Mg}{2}-absorbing CGM.  Here we discuss some of the weaknesses of
these assumptions and refinements to the model to be pursued in future work.

\subsubsection{The  Fiducial CGM and its
  Intrinsic Dispersion}

As described at the beginning of Section~\ref{sec.coherence}, a
fundamental assumption of our approach is the existence of a
``fiducial'' CGM -- i.e., that the \ion{Mg}{2} absorption
characteristics of all galaxies with a given set of intrinsic
properties (e.g., $M_*$, SFR) are equivalent.  Indeed, this assumption
is implicit in any study
using the assembly of single-sightline b/g probes to analyze CGM
absorption as a function of f/g host galaxy parameters. 
It is currently difficult to justify this simplification given the paucity of
systems for which absorption along more than one b/g sightline may be
analyzed.  This is of particular concern given that cosmological
``zoom'' simulations predict a large degree of variation in mass
outflow rates from a given galaxy on timescales $< 100$ Myr
\citep[although this effect is less pronounced at $z<1$; e.g.,][]{Muratov2015}.

The growing samples of gravitationally lensed QSO
sightlines passing close to bright f/g galaxies 
\citep[e.g.,][]{Chen2014,Zahedy2016} will eventually constrain the
dispersion in $W_{2796}$ within individual halos, and hence will
provide an independent estimate of our cosmic scatter parameter
$\sigma_{\rm C}$.  
 A yet more powerful experimental design is that of \citet{Lopez2018},
who used the bright, extended arc of a gravitationally lensed b/g
galaxy to study \ion{Mg}{2} absorption in a single f/g halo along 56
independent sightlines extending from $\mrperp = 15$ to 90 kpc.  The
arc, sampled at ${\sim}2-4$ kpc spatial resolution, revealed
significant scatter in $W_{2796}$ from sightline to sightline.
The magnitude of this scatter was reportedly smaller than that inferred
from QSO-galaxy pair samples \citep{Chen2010,Nielsen2013};
however, the authors caution that the latter included f/g galaxies with a broad
range of intrinsic properties (as well as absorption-selected
systems).  
A detailed comparison between such multi-sightline probes and
QSO-galaxy pair sightlines with similar f/g galaxy properties will likewise
provide highly valuable constraints on the intrinsic scatter of $W_{2796}$.
Finally,  advancements in UV-sensitive integral
field spectrographs (e.g., the Keck Cosmic Web Imager;
\citealt{Morrissey2018}) 
will soon make the direct detection of Ly$\alpha$ emission
from diffuse halo material routine, permitting constraints both on the
clumping scale of this emission in individual halos and comparisons of
emission properties among large galaxy samples.  While direct
detection of \ion{Mg}{2} in emission may be more challenging, it
will likely be possible for at least a handful of $z\sim0.3-1$ systems
\citep[e.g.,][]{Rubin2011,Martin2013}.

A second assumption important to our analysis is that there is no
evolution in this fiducial CGM between $z\sim 0.35-0.8$ (the epoch
probed by our PRIMUS pair sample) and the lower-redshift regime 
within which most of our QSO-galaxy comparison systems lie
($z\sim0.1-0.3$).  Stated more generally, our assumption is that both
our PRIMUS pairs and QSO-galaxy pairs probe equivalent circumgalactic
media as a function of $M_*$ and $\mrperp$.   One systematic bias to
consider is the decline in the average specific SFR (sSFR) for galaxies of a given
$M_*$ over this period \citep{Noeske2007,Speagle2014}, which could in
principle drive enhanced $W_{2796}$ values (or enhance variability
in $W_{2796}$ values) for higher-redshift systems.  In addition, the
masses of the dark matter halos hosting galaxies of a given $M_*$
decrease with cosmic time \citep[e.g.,][]{Moster2013}, which may
affect the observed velocity dispersion of cool clouds embedded in
these structures.  Finally, the virial radius of a dark matter
halo of a particular mass is redshift-dependent ($R_{\rm vir} \propto
(1+z)^{-2/3}$; \citealt{MB2004}),
such that our choice to compare CGM properties as a function of $\mrperp$
implies that the comparison is being carried out at inconsistent
values of $\mrperp / R_{\rm vir}$.

Regarding the first issue of sSFR evolution, \citetalias{GPG1}
demonstrated that there is a $+0.5$ dex offset between the median sSFR of
the star-forming f/g galaxies in the PRIMUS pair and the QSO-galaxy pair
samples consistent with the expected decline in star
formation activity from $z \sim 0.4$ to $z\sim 0.2$.  However, this
work also demonstrated consistency in the median $W_{2796}$ measured
around PRIMUS and QSO-galaxy subsamples of f/g galaxies selected to span the same
ranges in $M_*$.  Thus, at the sensitivity of our dataset, we find no
evidence for a systematic offset in $W_{2796}$ values due to
differences in star formation activity; however, we cannot rule out
the possibility of differing {\it dispersions} in $W_{2796}$ arising
from this evolution.  A sample of QSO-galaxy pairs with f/g host properties more
similar to those of the PRIMUS f/g galaxy sample will be needed to control
for this effect.

Regarding issues related to differences in the dark matter halo
masses and virial radii between these two samples, abundance matching
analyses predict that 
the halo mass for a central galaxy with $\log
M_{*}/M_{\odot} = 10$ decreases by  approximately $+0.2$ dex from $\log M_h /
M_{\odot} \approx 11.7$ to $\log M_h /
M_{\odot} \approx 11.5$  between
$z=0.5$ and $z=0.0$ \citep{Moster2013}.   
Accounting for the
difference in the median
redshifts of the two samples ($z=0.44$ vs.\ $z=0.25$), the corresponding change in halo virial velocity is only ${\sim}
20\mkms$ ($v_{\rm vir} \approx 117\mkms$ vs.\ $95\mkms$;
\citealt{MB2004}), implying a difference in the FWHM of the \ion{Mg}{2}
velocity distribution observed along the line of sight of $<40\mkms$.   
Any enhancement in $W_{2796}$ driven by
the larger velocity dispersion of the PRIMUS host halos will therefore likely be
$\lesssim0.2$ \AA\ \citep{Ellison2006}.
The difference in the virial radii of two
halos with $\log M_h /
M_{\odot} \approx 11.7$ and $\log M_h /
M_{\odot} \approx 11.5$ at these two epochs is also relatively 
insignificant: $R_{\rm vir}$ decreases from $\approx 157$ kpc at $z=0.44$ to $148$
kpc at $z=0.25$ (i.e., by only $\sim6\%$).  
Thus, we do not
consider evolution in host dark matter halo properties a major source of systematic uncertainty in our
modeling.

More generally, an intrinsic dependence of CGM properties on any host
galaxy property not included in our fiducial model (e.g., host galaxy
orientation and/or azimuthal angle; see Appendix~\ref{sec.azang}) will cause us to
overestimate the intrinsic scatter in $W_{2796}$.  Ultimately, as
mentioned above, significantly larger QSO-galaxy and galaxy-galaxy
pair samples will be required to reveal such detailed dependencies and
effectively mitigate
this issue.  We expect, however, that as long as (1) these intrinsic dependencies
are consistent between both pair samples and (2) the f/g galaxies in
these samples exhibit similar ranges in the relevant properties, the
results of our analysis are likely insensitive to such systematics.  



\subsubsection{Absorber and Background Beam Morphology}

Our modeling has also made some critical assumptions regarding
the morphology of \ion{Mg}{2} absorbers.  First, we have assumed that
every absorber has the same projected size regardless of its strength
($W_{2796}$).  However, the discussion presented in
Sections~\ref{subsubsec.lensing}  and \ref{subsubsec.photoionization}
motivates the exploration of models with stronger absorbers having
larger projected areas.  Given the small size of the PRIMUS pairs
dataset and its limited S/N, we do not attempt to incorporate such
refinements in the present work.  We also note that this type of
analysis will be best leveraged if detailed constraints on the UV
continuum morphology of the b/g galaxy sample are available.  {\it
  HST} imaging of these objects would enable a more precise prediction of
the observed absorber strength given a particular foreground
spatial distribution of $W_{2796}$.   Finally, future efforts must
certainly move beyond our invocation of square absorbers to
accommodate a more generic morphology as suggested by cosmological
simulations \citep[e.g.,][]{Faucher-Giguere2015,Nelson2016,
  Fielding2017,Oppenheimer2018}.




\subsection{Implications for the Thermodynamics of the CGM}\label{subsec.thermo}

Recent advancements in our empirical picture of 
the media surrounding galaxies have forced us to confront a
fundamental deficiency in our understanding of its governing physics.  On the one hand, 
numerous surveys have pointed to the ubiquity of cool ($T\sim10^4$ K),
photoionized material extending to $\mrperp > 100$ kpc from luminous
galaxies at low redshift
\citep[e.g.,][]{Chen2010,Werk2013,Stocke2013,Nielsen2013}.  On the
other hand, the frequent detection of \ion{O}{6} absorption along the
  same sightlines \citep[e.g.,][]{Tripp2008,Tumlinson2011,Savage2014},
  the detection in X-ray imaging of extended, hot ($T \sim 10^{6}$ K)
  halo material around nearby spirals \citep[e.g.,][]{Anderson2016},
  and a strong theoretical expectation of virialized infall
  \citep{ReesOstriker1977,WhiteRees1978,Keres2005} suggests
  that the CGM is also filled by a much hotter gaseous component.
 
 It is the physics permitting the coexistence of these two (or more)
  temperature components that is not understood.  The photoionization
  modeling analysis of \citet{Werk2014} 
  demonstrated that if the low-ionization absorption associated with
  the low-redshift CGM is assumed to arise
  from a single gas phase in ionization equilibrium at $T < 10^5~\rm K$,
  the implied volume density of this phase is two orders of magnitude too low to be
  in pressure equilibrium with a $T\sim10^6$ K hot medium. 
  One plausible resolution to this tension is the invocation of a
  range of densities for the cool phase (as in \citealt{Stern2016}).  
  Or, as discussed in \citet{Werk2014}, this apparent failure of an
  equilibrium solution may instead suggest that cool structures in the
  CGM are transient.

Indeed, there are several lines of evidence
  pointing in this latter direction.  First, the cool gas phase is
  frequently observed to trace
  large-scale gas outflows from 
  ${\sim} L_*$, star-forming galaxies at velocities  $\gtrsim200\mkms$
  \citep[e.g.,][]{Weiner2009,Martin2012,Rubin2013}.
  Second, inflows from the IGM or of
  previously-ejected gas are expected to
  persist to $z<1$ and involve
  material over a broad range of temperatures
  \citep[e.g.,][]{Keres2005,Nelson2013,Nelson2015,Ford2016}. 
  Third, recent theoretical works addressing this question posit that
  a CGM which is out of pressure equilibrium is a natural consequence
  of energy input from galactic outflows
  \citep{Thompson2016,Fielding2017}. 
In particular, the hydrodynamic
  simulations of CGM evolution discussed in \citet{Fielding2017}
  predict that the gas in dark matter halos less massive than
  $10^{11.5}~M_{\odot}$ never reaches hydrostatic equilibrium, and is
  instead supported by ram pressure and turbulence associated with
  galactic fountain flows.  
  Above this critical mass scale, the simulated gaseous halos are
   indeed supported by thermal pressure and 
feature a stable virial shock.
However, a cool phase
   arises from cold gas driven into the halo via feedback, from this
   cool wind seeding additional cooling, and/or  
   from (untriggered) thermal instability of the hot component.

   Moreover, many (perhaps all) of these processes involve the
   streaming of cool material through a surrounding hot medium, and
   hence will likely give rise to Rayleigh-Taylor or Kelvin-Helmholtz
   instabilities
   \citep{Agertz2007,Schaye2007,Crighton2015,Armillotta2016,Fielding2017}.  
These
   phenomena tend to disrupt the cool phase (in the absence of
   additional stabilizing mechanisms), and occur on physical scales
   which are far smaller than those typically resolved in the CGM of a
   halo-scale hydrodynamic simulation 
\citep{Nelson2015,Crighton2015,Fielding2017}.  
   It is therefore quite challenging to predict the overall contribution of
   these instabilities to the phase structure of halo gas; however, they
   would seem to make it yet more difficult to explain the
   coexistence of hot and cool media.

   Establishing the sizes of cool structures in the CGM is of primary
   importance to resolving this outstanding issue.  First of all, empirical
   constraints on the spatial extent of absorbing material greatly
   improve  estimates of (or limits on) the total masses of the
   structures.  
Second, the sizes and implied volume
   densities of these clouds are directly related to the timescales
   over which they will be disrupted by hydrodynamic instabilities.  
Given that non-equilibrium ionization conditions may be typical for
the CGM \citep[e.g.,][]{Oppenheimer2018}, such that 
size constraints from photoionization models are subject to
substantial systematic uncertainties, direct observation 
(using, e.g., extended b/g sources or lensed QSOs) is the most
promising method for assessing cloud  morphology.
We use the remainder of this section to discuss our constraints 
    on the mass and survival time of cool,
   \ion{Mg}{2}-absorbing circumgalactic structures.

\subsubsection{\ion{Mg}{2}-Absorbing Structure Mass}

   We have established a lower limit on the coherence length of
   \ion{Mg}{2} absorbers of $\ell_{\rm A} > \cllength$
   kpc.  As discussed in Section~\ref{subsubsec.interp_coherence}, our
   analysis requires that $W_{2796}$ remain approximately constant
   over this scale.  This requirement may be satisfied either by
   individual cool clouds that extend over $\ell_{\rm A} >
   \cllength$ kpc, or by a spatial correlation in the numbers and velocity
   dispersions of smaller clouds (respectively, scenarios (1) and (2)
   in Figure~\ref{fig.primus_discuss}).  Here we offer a calculation
   of the lower limit on the
   mass of these structures implied by our constraint on $\ell_{\rm
     A}$  under the assumption that they are best
   described by the former scenario.

As our \ion{Mg}{2} absorption lines are saturated and their velocity
components unresolved, we do not
   report column densities for these systems.  However, large $W_{2796}$
   values of ${>}0.5$ \AA\ (as are typical for our sample) 
commonly 
   arise from absorbers with column densities $N($\ion{Mg}{2}$) >
   10^{13.0}~\rm cm^{-2}$ (as shown in Section~\ref{subsubsec.degeneracies}).
   If we assume that \ion{Mg}{2} is the dominant ion in this gas phase
   \citep{Churchill2003,Narayanan2008}, adopt a solar abundance ratio
   ($\log \rm Mg/H = -4.42$; \citealt{SavageSembach1996}), and assume
   Mg is not depleted by dust (a very conservative assumption;
   \citealt{Jenkins2009}), our limit on the total hydrogen column
   density along these sightlines is $N(\rm H) > 10^{17.4}~\rm
   cm^{-2}$.
   Following \citet{Crighton2015}, we estimate  the total mass
   of each structure with the relation
 \begin{equation*}
  M_{\rm A} =   \ell_{\rm A}^3 n_{\rm H} \mu m_{\rm p},
\end{equation*}
with the hydrogen number density $n_{\rm H} = N(\mathrm{H}) /
\ell_{\rm A}$, and $\mu m_{\rm p}$ the mass per hydrogen atom with $\mu = 1.4$.  This
equation may be rewritten
\begin{equation*}
  M_{\rm A} = 1.1 \times 10^{4}~M_{\odot} \left (\frac{\ell_{\rm A}}{1.9~\mathrm{kpc}}\right )^2
  \left ( \frac{N(\mathrm{H})}{10^{17.4}~\mathrm{cm}^{-2}} \right ).
\end{equation*}
Each structure thus likely contains at a minimum ${\sim}10^4 M_{\odot}$
of material.  This mass is modest in the context of HVCs (which are
estimated to contain ${\sim}10^8 M_{\odot}$ including their ionized
component; \citealt{LehnerHowk2011}),  but is similar to the cool cloud mass estimated from the photoionization analysis
of CGM absorption at $z\sim2.5$ by \citet{Crighton2015}.

\subsubsection{Cool Cloud Survival}

 In the absence of stabilizing
mechanisms (e.g., magnetic fields; \citealt{McCourt2015}),
Kelvin-Helmholtz instabilities will destroy these clouds on a
timescale similar to the cloud crushing time expressed in Equation~\ref{eq.cct}.
Again assuming that each individual cloud extends over the length
scale $\ell_{\rm A}$,
we may rewrite this timescale as in \citet{Crighton2015}:
\begin{equation*}
  \tau \approx 19.6~\mathrm{Myr} \left ( \frac{\ell_{\rm A}}{\cllength~\mathrm{kpc}} \right )
  \left (\frac{v}{300\mkms} \right )^{-1} \left ( \frac{n_{\rm
        H}/n_{\rm halo}}{10}\right )^{1/2}
\end{equation*}
(see also \citealt{Agertz2007} and \citealt{Schaye2007}), with 
$v$ representing the relative velocity between the hot and cool phases, and 
$(n_{\rm  H}/n_{\rm halo})$ equal to the ratio of their number
densities.\footnote{The length scale used in this equation by \citet{Crighton2015} refers
to the radius of a spherical cloud.  
The factor required to adapt this relation to our assumption of
cubical clouds is of order unity, and we ignore it here.}
As noted in \S\ref{sec.intro}, this relation suggests that larger,
higher-density, slow-moving clouds will last longer in the hot halo
environment.

The density ratio $(n_{\rm  H}/n_{\rm halo})$ is the most poorly
constrained of the relevant quantities in this context; however, we may
refer to \citet{Werk2014} for guidance.  They demonstrated that for a
halo of mass $M_h=10^{12}\msun$, the 
virialized hot phase as modeled in \citet{MB2004} has $n_{\rm halo}
\approx 10^{-3}~\rm cm^{-3}$ within $\mrperp < 50$ kpc.  
This number density is very similar to their empirical estimate of
the density of the cool, photoionized phase.  We therefore contend
that potential 
variations in this ratio would likely lead to only a marginal decrease
in $\tau$.  

Regarding the velocity $v$, we expect this to depend on
the source of the material.  Large-scale outflows traced by
\ion{Mg}{2} absorbers similar in strength to those observed toward our
b/g sightlines occur ubiquitously among massive, star-forming galaxies
at $z\sim0.5$, and exhibit velocities of $\gtrsim 200-400\mkms$ 
\citep{Rubin2013}.  If these high velocities persist as the material
flows away from the galaxies, structures with 
$\ell_{\rm A} \sim \cllength$ kpc would be disrupted after $\lesssim 20$ Myr,
and after having traveled only ${\sim}6$ kpc.  
Structures as large as $\ell_{\rm A}\sim4$ kpc (i.e., similar in size to
the optical half-light radii of the host galaxy sample) moving at such high
speeds would survive for up to ${\sim} 41$ Myr, but would reach
distances of only $\mrperp = 13$ kpc in that time.  It is worth noting 
that unless there is a pervasive mechanism which acts to suppress 
hydrodynamic instabilities in galaxy halos of this size, the
\ion{Mg}{2} absorption observed to be moving at hundreds of kilometers
per second
when viewed down-the-barrel toward star-forming
systems  cannot also give rise to strong \ion{Mg}{2}
absorption at $\mrperp > 20$ kpc, even if the sizes of the absorbing
structures are well above our lower limit on $\ell_{\rm A}$.

Such stabilizing mechanisms have  by no means been ruled out by
empirical evidence.  On the contrary, magnetic fields are frequently
invoked to suppress cool cloud destruction
\citep[e.g.,][]{McClure-Griffiths2010,McCourt2015}, and are beginning to be recognized as a
common phenomenon in nearby spiral galaxy halos due to recent
advancements in radio continuum
surveys \citep[e.g.,][]{Krause2009,Wiegert2015}.
Analyses of high-resolution hydrodynamical simulations of the interaction of cool
clouds with a hot wind fluid suggest that radiative cooling and
thermal conduction can both inhibit
the mixing of the two phases
\citep{Mellema2002,Cooper2009,Armillotta2016}.  These mechanisms may
not always act in tandem, however; for instance, thermal conduction
may itself also be suppressed by magnetic fields \citep{McCourt2018}.
The internal structure of the clouds themselves may also aid in
prolonging their survival: as discussed in
\citet{McCourt2018}, such clouds may be prone to shattering into numerous overdense
fragments as they cool, which could in turn make them  more robust to
disruption than a monolithic structure
with uniform density.

On the other hand, the observed \ion{Mg}{2} absorbers may instead
originate in diffuse accretion streams and/or in gas that is stripped
from infalling satellite galaxies via tides or ram pressure
\citep[e.g.,][]{Stewart2011,Fumagalli2011,Stewart2013,Ho2017}.  Qualitative examination of
cosmological simulations predicting the presence of these inflows
suggests that they would indeed be coherent over scales of $\gtrsim 2$ kpc \citep{Stewart2011,Stewart2013},
satisfying our cloud size constraint.  Moreover, several streams which
overlap along the line of sight could yield the large observed values
of $W_{2796}$.  Yet another possible origin is the precipitation of
\ion{Mg}{2}-absorbing gas due to thermal instability of the hot
  virialized halo medium \citep{MB2004,Voit2015,Fielding2017}.  In either of
  these latter scenarios, we expect that the cool structures would be
  moving through their host halos at speeds comparable to the free-fall
  velocity.  For a halo with mass
  $M_{h} \sim 10^{11-12}~M_{\odot}$ at $z\sim1$, this velocity is predicted to be
  ${\sim}50-150\mkms$ \citep{GoerdtCeverino2015}.  From the equation
  for $\tau$ above, a cool structure with
  $\ell_{\rm A} \sim \cllength$ kpc and a velocity of $100\mkms$ would survive
  without disruption over ${\gtrsim} 60$ Myr.  As infalling streams may
   be significantly more extended than ${\sim}2$ kpc, their
  survival time may plausibly be much longer.

Ideally,  a hydrodynamic simulation of a forming galaxy would be used
to test these competing scenarios.  Indeed, insight gained from
simulations of individual cool structures interacting with an
ambient hot flow has grounded much of the foregoing discussion
\citep[e.g.,][]{Agertz2007,HeitschPutman2009,McCourt2015,Armillotta2016,McCourt2018}.
However, 
resolving the
hydrodynamic instabilities on the surfaces of these structures may be beyond the
capability of any current idealized or cosmological zoom
simulation.  \citet{Agertz2007} found that simulations using the adaptive mesh refinement (AMR)
technique require at least seven grid cells per cloud radius to
properly capture cold cloud evolution.  
Assuming the clouds have sizes near the limit of our
coherence constraint ($\ell_{\rm A} = \cllength$ kpc), 
this implies that cells ${<}140$ pc across are needed.
Recent AMR simulations do indeed have cells of approximately this
size, but only in the highest-density regions (i.e., in the galactic
disk; \citealt{Hummels2013,AgertzKravtsov2016}).  
Simulations making
use of smoothed particle
hydrodynamics (SPH) codes require at least 7000
particles per cloud in order to adequately resolve surface instabilities 
\citep{Crighton2015},  and would therefore need to adopt particle
masses ${<} 2\msun$ to resolve a structure of  $M_{\rm A} =
10^4~M_{\odot}$.  This limit is well below the baryonic particle
mass of state-of-the-art SPH simulations \citep[e.g.,][]{Muratov2017,Hopkins2017,Oppenheimer2018}. 
The shortfall in our current capability to resolve diffuse
structures in detail has led some to call for the implementation of
``sub-grid'' recipes to approximate the complex phase structure of the CGM
\citep[e.g.,][]{Crighton2015}.  Advancements in both simulation
techniques and in empirical characterization of these structures are
needed to ultimately reveal their physical origin, their lifetime, and
their role in feeding galactic star formation.

\section{Conclusions and Future Directions}\label{sec.conclude}

With the aim to characterize the small-scale structure of
the cool, photoionized gas phase of the circumgalactic medium (CGM), we have presented
a detailed analysis of \ion{Mg}{2} absorption measured along
sightlines to bright background (b/g) galaxies probing the environments
within projected distances $\mrperp < 50$ kpc of  $27$ star-forming foreground (f/g) galaxies
at redshifts $0.35 \lesssim z_{\rm f/g} \lesssim 0.80$ and having stellar masses $9.1 < \log
M_*/M_{\odot} < 11.1$.  The galaxy pair sample, first
described in \citetalias{GPG1}, was drawn from the PRIsm MUlti-object
Survey \citep[PRIMUS;][]{Coil2011primus}, and includes b/g galaxies as faint as
$B_{\rm AB} \lesssim 22.3$ over a redshift range $0.4 < z_{\rm b/g} < 1.3$.
Rest-frame near-UV spectroscopy of the sample from \citetalias{GPG1}
permits constraints on the \ion{Mg}{2}
$\lambda 2796$ absorption strength associated with the f/g
systems to a limiting equivalent width $W_{2796}\gtrsim0.5$ \AA\ in individual b/g
galaxy spectra.
Moreover, unlike the QSO sightlines typically used to probe
circumgalactic material in absorption, the b/g galaxies for which {\it
HST} imaging is available are spatially extended with
half-light radii $1.0~\mathrm{kpc} < R_{\rm eff} (z_{\rm f/g})< 7.9~\mathrm{kpc}$.

Our analysis also leverages a sample of $W_{2796}$ values measured in b/g QSO
sightline spectroscopy probing the halos of f/g galaxies with a similar range in
stellar mass ($M_*$) at $z\sim 0.1-0.3$ from \citet{Chen2010}
and \citet{Werk2013}.
By making the assumptions that (1) the \ion{Mg}{2}-absorbing CGM
exhibits the same $W_{2796}$ distribution for every host
galaxy of a given $M_*$ as a function of $\mrperp$; and that (2) the quantity
$\log W_{2796}$ has a Gaussian distribution  with a constant
dispersion 
 (i.e., with a dispersion that does not depend on $M_*$ or $\mrperp$), we use this QSO-galaxy
 pair sample to construct a ``fiducial''
  model for the $\log W_{2796}$ distribution in the CGM of
 low-redshift galaxies.  

Then, by adopting the assumption that all
 \ion{Mg}{2} absorbers have the same projected surface area
 ($A_{\rm A}$) regardless of
 their strength, we use this fiducial model show how the $\log W_{2796}$
 distribution observed along a given set of sightlines 
depends on the ratio of the surface area of the b/g
 probes  ($A_{\rm G}$) to that of the absorbers ($x_{\rm A} \equiv
 A_{\rm G}/A_{\rm A}$).
We compare these model
 distributions to the sample of $\log W_{2796}$ values
  measured toward PRIMUS b/g galaxy sightlines, 
rejecting the null hypothesis that the observed and modeled distributions
are drawn from the same parent population for values of the ratio
$x_{\rm A} \ge 15$ 
at $95\%$ confidence.  This limit, in
 combination with the observed distribution of b/g galaxy sizes,
 requires that the coherence scale of \ion{Mg}{2}
 absorption -- that is, the length scale over which $W_{2796}$ does
 not vary -- is $\ell_{\rm A} > \cllength$ kpc.  This is the first such constraint
 on the morphology of cool, photoionized structures in the inner CGM
 (within $\mrperp < 50$ kpc)
 of ${\sim} L^*$ galaxies at $z<1$.

While this limit does not necessarily imply that the absorbing
material occupies structures extending over this length scale, it
requires that regions of the CGM giving rise to a particular observed
$W_{2796}$ are spatially correlated.  Complementary experiments using
gravitationally lensed QSO sightlines to probe strong \ion{Mg}{2}
absorbers (that are similar in strength to those observed in the inner CGM)
likewise suggest that these systems exhibit similar velocity structures
and absorption strengths over scales ${\gtrsim} 1-5$ kpc \citep[e.g.,][]{Monier1998,Rauch2002,Chen2014}.  
And although photoionization modeling of metal-rich absorbers 
identified in $z\sim2$ galaxy halos
tends to
yield smaller cloud sizes ($\lesssim 0.5$ kpc; e.g., 
\citealt{Crighton2015,Lau2016}), such modeling of circumgalactic
absorption in the low-redshift universe is suggestive of more extended
structures consistent with our coherence limit \citep{Werk2014}.


The primary factors limiting the strength of our constraints 
are the small sizes of the QSO-galaxy pair and projected galaxy pair
samples and the relatively low S/N of our b/g galaxy spectroscopy.  
Within the next few years, however, ongoing and prospective
large-scale surveys promise dramatic improvements in these sample
sizes.  The SDSS-IV/eBOSS program will nearly quadruple the 
surface density of known QSOs at $0 < z < 1.6$ relative to that
uncovered by earlier phases of the SDSS by 2020
\citep{Dawson2016,Abolfathi2017,LanMo2018}.  Beginning in 2018, the Dark Energy
Spectroscopic Instrument (DESI) will pursue
spectroscopy over 14,000 deg$^2$ of sky, obtaining samples of
${\sim}100$  QSOs
per deg$^2$ at $1<z<2$ and ${\sim}725$ emission-line-selected galaxies
per deg$^2$ with $0.6 < z <
1.0$  \citep{DESI2016}.  This unprecedented dataset will
therefore yield over 20,000 projected pairs of these objects with angular
separations $< 10\arcsec$, 
effectively increasing the sample size of QSO-galaxy pairs included in
the present work by more than two orders of magnitude.

Significant expansion of the projected galaxy pair sample will be more costly, 
but nonetheless will also be within reach over the next few years.  A
crucial component will be the identification of numerous additional galaxies which are
sufficiently bright to serve as effective background light sources
within reasonable integration times
with current facilities (i.e., Keck/LRIS and VLT/FORS2).  
The ongoing Dark Energy Survey (DES; \citealt{Drlica-Wagner2017}) and Dark Energy Camera Legacy
Survey (DECaLS\footnote{http://legacysurvey.org/decamls}) will permit
the photometric selection of galaxy pair candidates over 14,000
deg$^2$ to a limiting magnitude of $g\lesssim24.0$.  The surface
density of galaxy pairs with $\mrperp < 50$ kpc and with b/g galaxies
having $B_{\rm AB} <
23.3$ identified in the PRIMUS survey is ${\sim} 9~\rm deg^{-2}$,
implying that at least 100,000 such targets could be discovered within
the DES and DECaLS footprints.  With a focus on those pairs with the
brightest b/g objects, the present sample of ${\sim}30$ pairs could be
tripled within a modest Keck/LRIS allocation \citep{Lee2014obs}.  
Such improvements would add substantial leverage to constraints on
models invoking a more realistic (and complex) relationship between absorber
morphology and strength than that
explored here.

Moreover, 30m-class ground-based telescopes will enable high-S/N
spectroscopy of a significantly fainter galaxy population.  
With the Large Synoptic Survey Telescope poised to provide extremely deep,
 multiband imaging over 18,000 deg$^2$ \citep{Ivezic2008}, and 
because such
galaxies have a much higher sky density than QSOs, this technique will
ultimately become
the primary method used to probe the $z\gtrsim2$ CGM
 \citep{Steidel2009}.  
It therefore behooves us to carefully assess the effects of using
extended b/g sources on the relevant observables (i.e., equivalent
width, column density, velocity structure, etc.) by pursuing detailed
comparisons with pencil-beam surveys.  These efforts will both
optimize future experiments, and offer novel constraints on
the small-scale structure of the cool phase of the gas that pervades galactic environments.







\acknowledgements
K.H.R.R acknowledges support for this project from the Clay
Postdoctoral Fellowship.  A.L.C. acknowledges support from 
NSF CAREER award AST-1055081.

These findings are in part based on observations collected at the
European Organisation for Astronomical Research in the Southern
Hemisphere under ESO programs 088.A-0529(A) and 090.A-0485(A).
In addition, a significant portion 
of the data presented herein were obtained at the W. M. Keck
Observatory, 
which is operated as a scientific partnership among the 
California Institute of Technology, the University of California, and
the National Aeronautics and Space Administration. The Observatory was
made possible by the 
generous financial support of the W. M. Keck Foundation.

It is our pleasure to thank Brice M{\'e}nard, Nicolas Tejos, Jessica
Werk, Andrew Fox, James Bullock, and Nicolas Bouch{\'e} for enlightening discussions
which improved this analysis.  We gratefully acknowledge
J. X. Prochaska for his reading of this manuscript, and for sharing
valuable insight on this topic.  We also thank the anonymous
referee, whose suggestions helped to improve this work.

Finally, the authors wish to recognize and acknowledge the very
significant cultural role and reverence that the summit of Mauna Kea
has always had within the indigenous Hawaiian community. We are most
fortunate to have the opportunity to conduct observations from this
mountain.

\bibliography{adssample}

\appendix

\section{The Significance of Foreground Galaxy Orientation}\label{sec.azang}

The analysis presented in Section~\ref{sec.coherence}
rests on the assumption that the
fluctuation of $W_{2796}$ within a halo hosting a galaxy of a
particular $M_*$ at a given $\mrperp$ is independent of all other
intrinsic galaxy properties.  However, several studies now suggest
that $W_{2796}$ may depend to some degree on the placement
of the b/g sightline relative to the orientation of the f/g host's stellar disk.  
This issue was first addressed by
\citet{Bordoloi2011}, who studied \ion{Mg}{2} absorption in stacked b/g galaxy
sightlines probing the halos of f/g edge-on disk-dominated galaxies.
These authors subdivided their sample by the azimuthal angle ($\Phi$)
of the b/g sightline; i.e., the angle between the f/g disk major axis and
the location of the b/g galaxy on the sky (taking the center of the
f/g galaxy to be the origin).  They found that in sightlines with 
$\Phi < 45^{\circ}$ (located along the minor axis of the
f/g disk) and with $\mrperp < 40$ kpc, the median EW of the blended
\ion{Mg}{2} doublet was
enhanced by ${\sim}0.8$ \AA\ relative to sightlines with $\Phi >
45^{\circ}$.  This enhancement was interpreted as a signature of bipolar
galactic winds.  

\citet{Bouche2012} and \citet{Kacprzak2012} also
investigated this issue using a sample of projected QSO-galaxy pairs, 
the latter reporting a $20-30\%$ enhancement in the covering fraction
of $W_{2796} > 0.1$ \AA\ absorption close to both the minor and major
axes of star-forming galaxies, and finding that the strongest
absorbers were detected in sightlines having $\Phi$ within $< 50^{\circ}$ of the minor axis
at $\mrperp < 40$ kpc.
More recently, \citet{Lan2014} assessed the numbers of
edge-on star-forming galaxies within $\mrperp < 50$ kpc of SDSS QSO
sightlines exhibiting $W_{2796} > 1.5$ \AA\ absorbers, finding that
there were more of these galaxies oriented such that the
corresponding QSO sightline probed their minor axes.  Taken together,
these measurements  suggest that the incidence of the
strongest \ion{Mg}{2} systems depends on the orientation of the b/g
sightline relative to the f/g galaxy's disk.

We therefore consider here whether this dependence should be
included in our fiducial model for the \ion{Mg}{2}-absorbing CGM.  
To determine whether a minor-axis enhancement of $W_{2796}$ is exhibited 
by our QSO-galaxy comparison sample, we make use of the photometry of
the f/g galaxies reported in SDSS data release
10\footnote{http://cas.sdss.org/dr10/en/home.aspx} \citep{Ahn2014}.
We use the results
of the fits of
an exponential disk model to the SDSS $r$-band imaging of these
galaxies as indicators of the disk
axis ratios ($b/a$) and position angles.  
Of the 50 QSO-galaxy pairs with $\mrperp < 50$ kpc and with
star-forming f/g hosts in this sample, SDSS DR10 photometry is
available for 48.
We adopt the simple
assumption that the inclination of each galaxy is given by $i =
\arccos(b/a)$, and calculate $\Phi$ from the angle between each disk
position angle and the corresponding QSO coordinate.  Note that this
reference frame is different from that used in \citet{Bordoloi2011},
in that low values of $\Phi$ indicate sightlines along the disk major axes.

\begin{figure}
\begin{center}
\includegraphics[angle=0,width=0.5\columnwidth,trim=0 20 0
0,clip=]{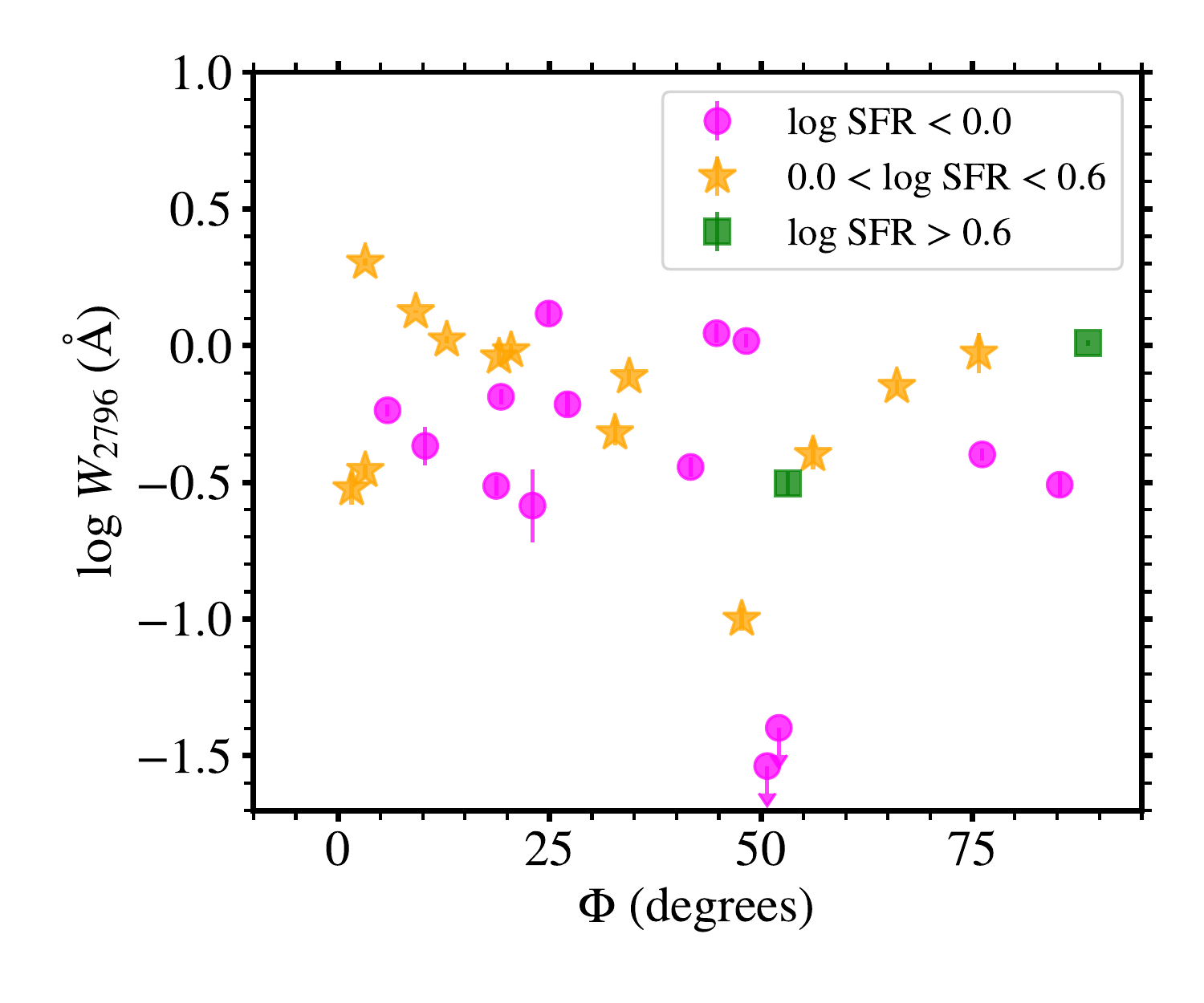}
\caption{$\log W_{2796}$ versus the azimuthal angle of the b/g sightline
  for star-forming f/g hosts within $\mrperp < 50$ kpc and having
  inclinations $i > 50^{\circ}$ in our
  QSO-galaxy comparison sample.  Pairs with f/g galaxies having low,
  intermediate, and high values of SFR are indicated with magenta
  circles, orange stars, and green squares.  Sightlines with
  $\Phi \sim 0^{\circ}$ and ${\sim} 90^{\circ}$ are close to the major
  axis and minor axis, respectively.
\label{fig.primus_azang}}
\end{center}
\end{figure}

We show the distribution of $\log W_{2796}$ versus $\Phi$ for all
star-forming f/g galaxies having inclinations $i > 50^{\circ}$ in
Figure~\ref{fig.primus_azang}, with the symbols color-coded by SFR as
in Figure~\ref{fig.primus_qsofid}.  
Only three of the 13 systems with $\Phi > 40^{\circ}$ have 
$W_{2796} > 1.0$ \AA, while
four of the 16 systems with $\Phi < 40^{\circ}$
meet this criterion.  
Indeed, we see no evidence for a
significantly higher incidence of strong absorbers at high $\Phi$ in
this sample.

There are several important differences between the experimental designs of the studies
mentioned above and that of the analysis shown in
Figure~\ref{fig.primus_azang} that may give rise to the apparent
inconsistency of these results.
For instance, the \citet{Bordoloi2011} study 
focused on f/g hosts at $z\sim0.7$, which are likely to have higher
SFRs than the f/g systems in our QSO-galaxy pair  sample and hence may drive
stronger large-scale winds, leading to a stronger enhancement
in bipolar \ion{Mg}{2} absorption.  
Our sample may simply be too small to reveal a significant
  dependence of
  $W_{2796}$ on $\Phi$: whereas the \citet{Bordoloi2011} work included
  54 pairs with edge-on f/g galaxies, with projected separations $\mrperp < 50$ kpc, and with QSO sightlines passing
  within $45^{\circ}$ of the disk minor axis,
  our QSO-galaxy pair sample contains only ${\sim}11$ systems
  with these properties.
Moreover, the \citet{Bouche2012} study included only QSO-galaxy pairs
  exhibiting strong \ion{Mg}{2} absorption, and could not assess the
  azimuthal angle distribution of weak absorbers.
The f/g galaxies 
discussed in \citet{Kacprzak2012} have
a wide range in redshifts ($0.1 \lesssim z \lesssim 1.1$); in
addition, about ${\sim}50\%$ of their f/g sample is \ion{Mg}{2}
absorption-selected, and hence may be biased to exhibit larger
$W_{2796}$ overall.  

In any case, a larger 
sample of high-S/N QSO sightline spectroscopy probing f/g galaxy halos at $\mrperp < 50$
kpc and selected without 
regard for the halo absorption strength
is needed before we may carry out a sensitive test of the
azimuthal angle dependence of $W_{2796}$ at $z\sim0.2$.  Given this
state of affairs, we conclude that the inclusion of such a dependence
for the $W_{2796}$ distribution in our fiducial CGM model is unjustified.

\end{document}